# Demonstration of a Polarization-Agnostic Geometric Phase in Nonlocal Metasurfaces

Adam Overvig[1,*], Yoshiaki Kasahara[1,2,*], Gengyu Xu[1], and Andrea Alù[1,3,†]

[1]Photonics Initiative, Advanced Science Research Center at the Graduate Center of the City University of New York

[2]Department of Electrical and Computer Engineering, the University of Texas at Austin

[3]Physics Program, Graduate Center, City University of New York

* These authors contributed equally. †Corresponding author: aalu@gc.cuny.edu

## Abstract

*Symmetry-driven phenomena arising in nonlocal metasurfaces supporting quasi-bound states in the continuum (q-BICs) have been opening new avenues to tailor enhanced light-matter interactions via perturbative design principles. Geometric phase concepts—observed in many physical systems—are particularly useful in nonlocal metasurfaces, as they enable to locally pattern the q-BIC scattering rate and phase across the metasurface aperture without affecting the delocalized nature of the q-BIC resonance. However, this control typically comes with stringent limitations in terms of efficiency and/or of polarization operation. Here, we unveil a new form of geometric phase control, accumulated along a continuous contour of geometric perturbations that parametrically encircle a singularity associated with a bound state. This response is obtained regardless of the chosen polarization state, which can be rationally specified by a judicious choice of the perturbation geometry. Our findings extend geometric phase concepts to arbitrary polarizations, including linear and elliptical states, with near-unity scattering efficiency. This polarization-agnostic geometric phase offers new opportunities for wavefront manipulation, multifunctional operation and light emission, with applications in*

*augmented reality, secure communications, wireless systems, imaging, lasing, nonlinear and quantum optics.*

**Introduction**

Quasi-bound states in the continuum (q-BICs) are long-lived optical states controlled by symmetry-breaking perturbations applied to engineered thin films. Q-BICs are characterized by the property that their quality (Q-) factor varies inversely proportionally to the square of the magnitude of the perturbation [1]. Their strong dependence on angle, associated with long in-plane propagation lengths correlating light across the aperture [2], makes these metasurfaces spatially *nonlocal* [3]. Recent efforts have been leveraging their capability to precisely engineer the optical response as a function of incident momentum [4] for volumetric imaging [5], analog optical computation [6],[7], image differentiation [8]-[10], squeezing free-space [11]-[12], sensing [13], and controlling the coherence of thermal emission [14]. While excellent at tailoring the impinging wavefront in momentum space, these implementations have a limited control over real space features, due to the extended nature of the underlying q-BIC resonance.

This functionality is in stark contrast to *local* metasurfaces, which can be engineered to shape the impinging light by locally patterning their nanostructures in real space [15]-[17]. The most common approach to engineer the wavefront using local metasurfaces is based on tuning the nanostructure resonance across the metasurface aperture, however this approach suffers from efficiency limitations and nonidealities, since their geometry discontinuously jumps from large to small as the phase wraps around $2\pi$ [16]. Geometric phase engineering [18]-[20], e.g., exploiting the Pancharatnam-Berry (PB) phase, overcomes this constraint. Indeed, while limited to circularly polarized (CP) waves, the PB phase does not suffer from geometric discontinuities when imparting smoothly varying phase profiles across the metasurface aperture, a stark advantage stemming from its topological nature [21]. Metasurfaces operating in a gray area between local and nonlocal responses have been recently explored for angle-dependent wavefront shaping, such as bianisotropic [22]-[23], Huygens' [24]-[25], asymmetric [26], supercell [27],[28], and metagrating [29]-[30] approaches. Both local and quasi-local metasurfaces, however, suffer from limited spectral control: for instance, the response away from the design frequency is generally suboptimal and poorly controllable. High-Q anomalous deflection is also possible with perturbative metagrating approaches [31] and reflectarrays [32],[33], but these approaches also suffer from limited efficiency, and lack of flexibility and polarization control.

A new class of metasurfaces, dubbed *diffractive nonlocal metasurfaces* [34]-[40], takes the best features from local and nonlocal metasurfaces. They are composed of highly symmetric arrays of carefully perturbed unit cells, tailored to exclusively manipulate delocalized resonant states with minimal scattering off resonance. Selection rules [35] governing the excitation of these states have enriched q-BIC approaches with vectorial and phase properties in single-layer arrays with controllable direction of linear polarization (LP) [36],[41], and in twisted stacked layers supporting chiral responses [37],[42],[43]. By carefully perturbing these properties in space, it is possible to achieve manipulation [38] and generation [39],[41] of custom wavefronts with extreme selectivity in both frequency and wavefront shape, without affecting the extended nature of their nonlocal q-BIC resonance. Preserving the nonlocal nature of the mode requires careful choice over the perturbations that locally shape the metasurface response, hence structural discontinuities associated with phase wrapping need to be avoided for operation with near-unity efficiency [44]. Hence, spatial patterning of diffractive nonlocal metasurfaces has so far relied on geometric phase approaches, with demonstrations limited to operation with circular polarization, and/or with limited efficiency. A new form of geometric phase with polarization-agnostic features is highly desirable to extend the capabilities of diffractive nonlocal metasurfaces to custom polarization states with high efficiency.

Here, we introduce a general perturbation scheme capable of simultaneous and independent control over the Q-factor ($Q$), phase ($\Phi$) and polarization state (definable by the elliptical parameters $\psi, \chi$) of q-BICs supported by nonlocal metasurfaces. This approach offers, for the first time, full control over all four degrees of freedom (DoF) intrinsic to a Fano resonance ($Q, \Phi, \psi, \chi$), and in a manner individually addressable by each meta-unit of a nonlocal metasurface, i.e., enabling $Q(x,y), \Phi(x,y), \psi(x,y), \chi(x,y)$ for a metasurface aperture in the $(x,y)$ plane. The key is a geometric phase that is polarization-agnostic by construction, which we name the *decay phase*. Based on an all-dielectric platform for operation in the near infrared regime, we demonstrate linearly polarized (LP) light with a geometric phase by tuning the relative scattering strength of the top and bottom interfaces of a nonlocal metasurface, and show that this phenomenon is associated with parametrically encircling a topological singularity corresponding to an artificial bound state in the continuum (a-BIC). By simultaneously tuning the relative scattering strength while rotating the LP state of each interface extends this geometric phase to arbitrarily elliptical polarized (EP) states, enabling nonlocal phase gradients selective to an arbitrary polarization of choice, and with no geometrical discontinuities, and resulting in highly efficient phase manipulation for any custom polarization. Leveraging the inherent universality of phenomena emerging from symmetry, we experimentally demonstrate nonlocal phase gradients selective to LP and EP light in metasurfaces operated at millimeter-waves [45].

**Available degrees of freedom**

We begin by considering the DoF of nonlocal metasurfaces excited at normal incidence. In this work, we are interested in devices supporting a broadband (background) response with near-unity transmission, punctuated by a narrowband Fano resonance that reflects with a Jones matrix of the form (see Supplementary Materials)

$$J_r = |e^*_1\rangle\langle e_1|, \qquad (1)$$

where $|e_1\rangle$ has 4 DoF. Ignoring losses, these DoF correspond to the set $(Q, \Phi, \psi, \chi)$. In this work, we introduce a scheme that offers rational control over all these four DoF. Simultaneous control over the nonradiative scattering rate, e.g., by selectively doping silicon at near infrared frequencies, would readily extend this technique to control both Q-factor and peak reflectance; here we consider the lossless scenario, wherein the peak reflectance of a meta-unit is fixed at unity [46].

For comparison, in Fig. 1(a) we sketch a conventional q-BIC approach for nonlocal metasurfaces [36], in which elliptical inclusions (refractive index $n = 1.45$, matching the superstrate and substrate) form a dimerized unit cell with dimensions $a \times 2a$, arrayed in a thin film ($n = 3.45$ and height $H = 0.5\mu m$), where $a = 400nm$ is the array pitch. A perturbation $\delta_1$ quantifies the deviation of the ellipses from perfect circles; their orientation angles are $\alpha_1$ and $\alpha_1 + 90°$. For small perturbations, the q-BIC scatters free-space radiation with $Q \propto 1/\delta_1^2$ and a LP angle $\psi \approx 2\alpha_1$ [36]. This dependence follows the selection rules for q-BICs [35] derived by considering end-fire coupling at *each* interface of the thin film. Here, the out-of-plane symmetry of the device enforces identical selection rules at the top and bottom interfaces [Fig. 1(b)], limiting the q-BIC to LP responses at normal incidence. Rotating this polarization spatially by varying $\alpha_1$ enables wavefront shaping when CP light is incident, but the achiral response splits the energy in four ways [Fig. 1(c)], capping the anomalous reflection efficiency at 25% [35],[36].

To overcome this limitation, the scattering at the top and bottom interfaces must be distinct [34], implying a stack of two perturbation layers at minimum. Ref. [37] added the orientation angle of the top layer $\alpha_2$ as an additional DoF [Fig. 1(d)], yielding distinct LP states at each interface [Fig. 1(e)] and hence arbitrary control over the resonant polarization state $(\psi, \chi)$. In the specific case of CP, a geometric phase enables anomalous reflection with near-unity efficiency for the chosen CP [Fig 1(f)]. We note that this geometric phase is associated with an infinite degeneracy in constructing CP light out of the two arbitrarily oriented LP states of each interface, here tailored by the metasurface parameters $(\alpha_1, \alpha_2)$. Unfortunately, the parameters $(\delta_1, \alpha_1, \alpha_2)$ afford only a single implementation for a given EP state, meaning that the observation of geometric phase is limited to CP in this system, as expected.

To overcome this important limitation, here we introduce a fourth geometric DoF: the ellipticity of the top layer with perturbation $\delta_2$, distinct from the bottom layer perturbation $\delta_1$ [Fig. 1(g)]. At first glance, changing $\delta_2$ appears to simply change the Q-factor by altering the overall perturbation strength. Surprisingly, however, altering the relative strength of the two scattering events [Fig. 1(h)] introduces a continuous degeneracy of EP states. As in the CP scenario, this infinite degeneracy is ultimately discriminated by the phase of the associated scattering process — the *decay phase* — enabling selective anomalous reflection of arbitrary EP states with unitary efficiency [Fig. 1(i)].

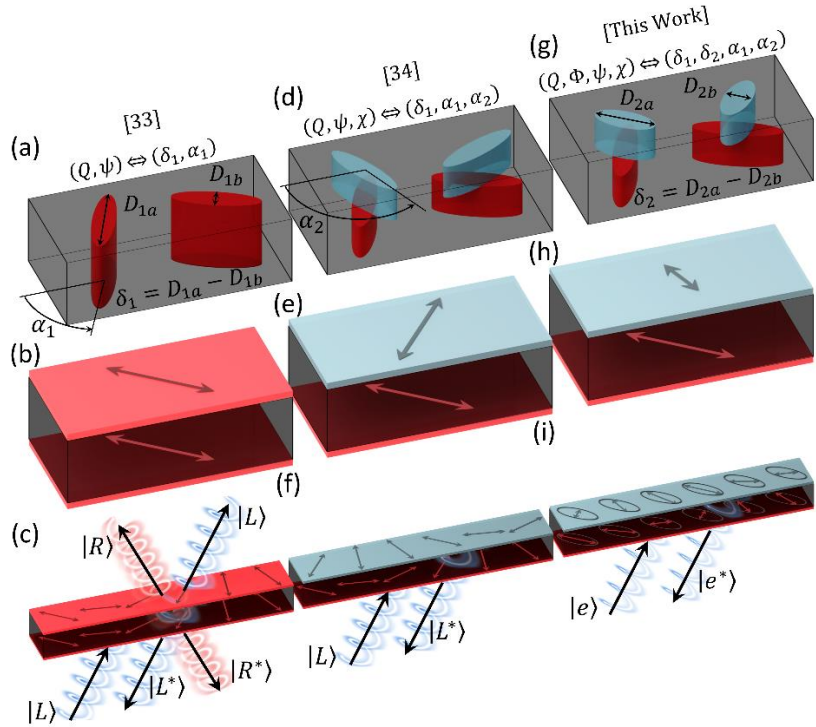

**Figure 1**. Q-BICs with progressively increasing control over the four DoF $(Q, \Phi, \psi, \chi)$. (a) Devices based on the approach in [36] may control $(Q, \psi)$ by manipulating the magnitude of the perturbation ($\delta_1$, determining the ellipticity of the inclusions) and orientation angles of the perturbed ellipses ($\alpha_1$, where the second ellipse is oriented at $\alpha_1 + 90°$). (b) Characteristic dipole moments at each interface induced by the scheme in (a). (c) Characteristic wavelength-exclusive wavefront shaping functionalities enabled by the scheme in (a). (d-f) Same as (a-c), but based on the approach in [37], which controls $(Q, \psi, \chi)$ by adding separate control over a top layer (blue), composed of identical ellipses as the bottom layer (red) but with orientation angle $\alpha_2$. (g-i) Same as (a-c), but based on the approach introduced here, controlling $(Q, \Phi, \psi, \chi)$ by adding a second magnitude of perturbation $\delta_2$ determining the ellipticity of the top layer inclusions.

## The decay phase: a polarization-agnostic geometric phase

In our analysis, we first consider LP light, demonstrating that our geometric phase can control with unitary efficiency LP waves by carefully perturbing the top and bottom layers [Fig. 2(a)], and then we discuss general EP states. Figure 2(b) shows full-wave simulations of the resonantly reflected phase and Q-factor as a function of $(\delta_1, \delta_2)$ when $\alpha_1 = \alpha_2 = 0$ and *x* polarized light is normally incident from the substrate. We observe a hallmark of the geometric phase [19]: a topological feature, here of order two, associated with parametrically encircling a singularity, here at $(0,0)$. Indeed, the perturbation vanishes at the origin, and the mode [Fig. 2(c)] is bound due to the addition of a discrete translational symmetry with periodicity $a \times a$. This mode is bound within the 'artificial continuum' associated with the periodicity $a \times 2a$: it is an 'a-BIC' [47]. It is only in the presence of the period doubling perturbation that the state is folded into the continuum [Fig. 2(d)], where it is promoted to a q-BIC endowed with properties exclusively imparted by the geometric perturbation. The spectral reflection and phase (computed by full-wave simulations) of four example meta-units are shown in Fig. 2(e) and 2(f), respectively. Crucially, the contour of devices in Figs. 2(a,b) is a closed loop of geometrical parameters, arriving back to the original geometry after accumulating a phase $\Phi = 4\pi$. Changing variables to $\delta_1 = \delta_0 \cos(\delta)$ and $\delta_2 = \delta_0 \sin(\delta)$, where $\delta_0$ is the radius and $\delta$ is the angular location along the contour, we find the approximate relations $Q \propto 1/\delta_0^2$ and $\Phi \approx 2\delta$.

To justify these relations, we develop a temporal coupled mode theory (TCMT) [48] for this system, demonstrating that the q-BIC decays with coefficients $|d_{up}\rangle$ towards the superstrate and $|d_{down}\rangle$ towards the substrate obeying (see Supplementary Materials)

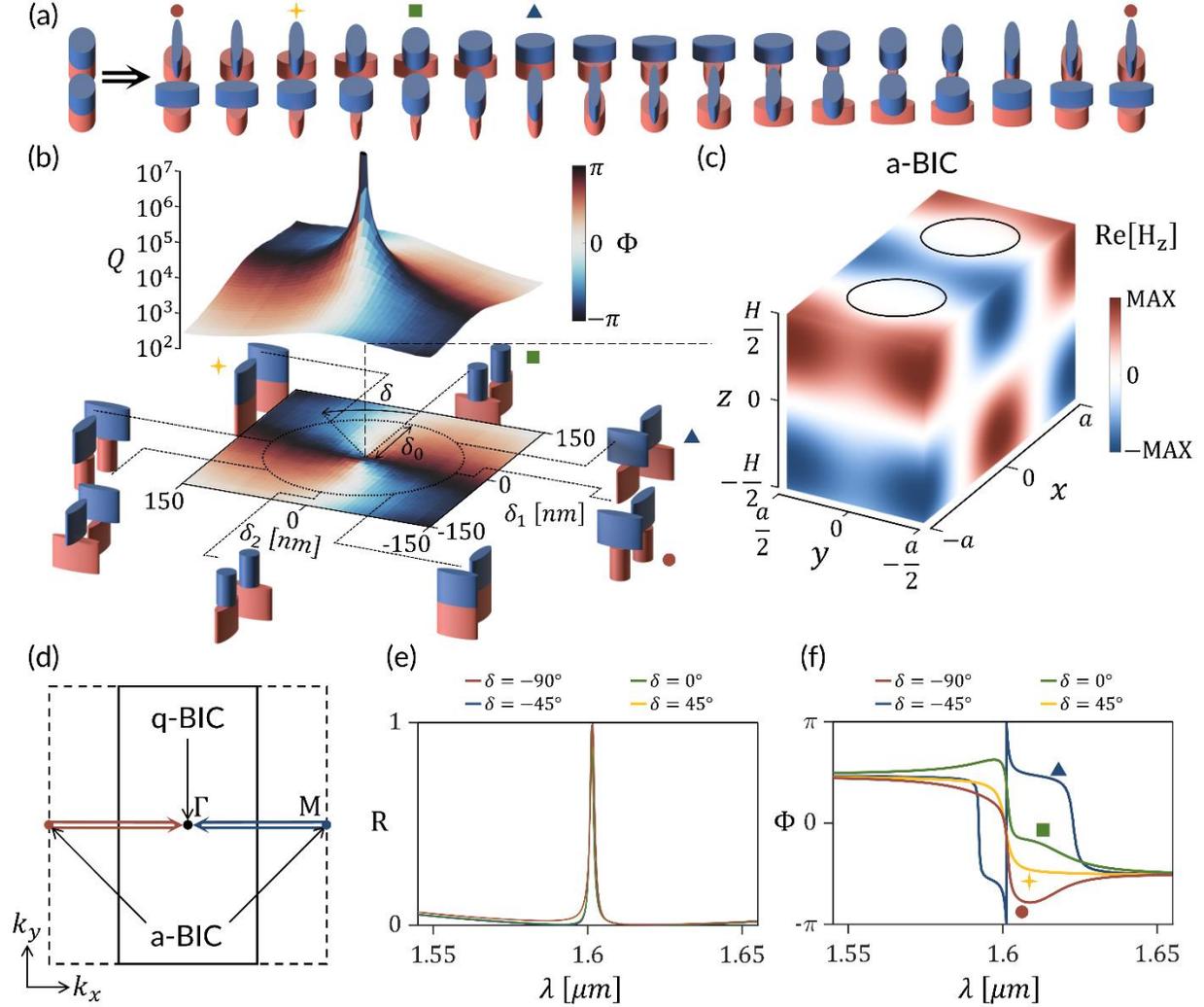

**Fig. 2**. **Resonantly reflected geometric phase for linear polarization.** (a) Dimerizing perturbation scheme with a geometric phase for x-polarized light. (b) Q-factor, $Q$, and reflection phase, $\Phi$, at resonance for devices with varying perturbation strengths $\delta_1$ and $\delta_2$. Eight meta-unit geometries are highlighted along the dashed contour, which encircles a topological feature of order two. (c) Field profile of the a-BIC corresponding to the singularity in (b). (d) Zone-folding diagram relating the a-BIC to the q-BICs upon perturbation. (e) Reflection and (f) phase profiles for meta-units corresponding to the markers shown in (a,b).

$$|d_{up}\rangle \propto i\delta_0 \left\{ \cos(\delta) \begin{bmatrix} \cos(2\alpha_1) \\ \sin(2\alpha_1) \end{bmatrix} - i\sin(\delta) \begin{bmatrix} \cos(2\alpha_2) \\ \sin(2\alpha_2) \end{bmatrix} \right\}$$
$$|d_{down}\rangle \propto i\delta_0 \left\{ -i\cos(\delta) \begin{bmatrix} \cos(2\alpha_1) \\ \sin(2\alpha_1) \end{bmatrix} + \sin(\delta) \begin{bmatrix} \cos(2\alpha_2) \\ \sin(2\alpha_2) \end{bmatrix} \right\}$$
(2)

When light is incident from the superstrate (substrate), then in Eq. (1) we have $|e_1\rangle \propto |d_{up}^*\rangle$ ( $|e_1\rangle \propto |d_{down}^*\rangle$ ), and our TCMT provides a closed form expression for the Jones matrix associated with the scattering event. The factor $-i$ in front of the $\sin(\delta)$ term in $|d_{up}\rangle$ is consistent with the scattering features of symmetric thin films: for a real-valued reflection Fresnel coefficient $r$, the corresponding transmission coefficient is $-it$, where $t$ is real-valued [48]. Correspondingly, light coupling to the superstrate from the bottom interface ['indirect' scattering, controlled by $(\delta_1, \alpha_1)$] has a phase factor $-i$ relative to the coupling to the superstrate from the top interface ['direct' scattering, controlled by $(\delta_2, \alpha_2)$]. An analogous property holds for coupling to the substrate. In other words, relative to some reference phase, one interface controls the real part of the scattering, while the other interface controls its imaginary part. When $\alpha_1 = \alpha_2 = 0$, it is simple to recover the relations $Q \propto 1/\delta_0^2$ and $\Phi = 2\delta$, and we see that moving along the contour in Fig. 2(b) varies the phase by weighting the magnitudes of the direct and indirect scattering components. With the resulting TCMT, we may analytically map the set of geometrical parameters $(\delta_0, \delta, \alpha_1, \alpha_2)$ to and from $(Q_0, \Phi, \psi, \chi)$, as detailed in the Supplementary Materials, populating a library of meta-units for completely and independently control all four available DoF via a set of analytical equations.

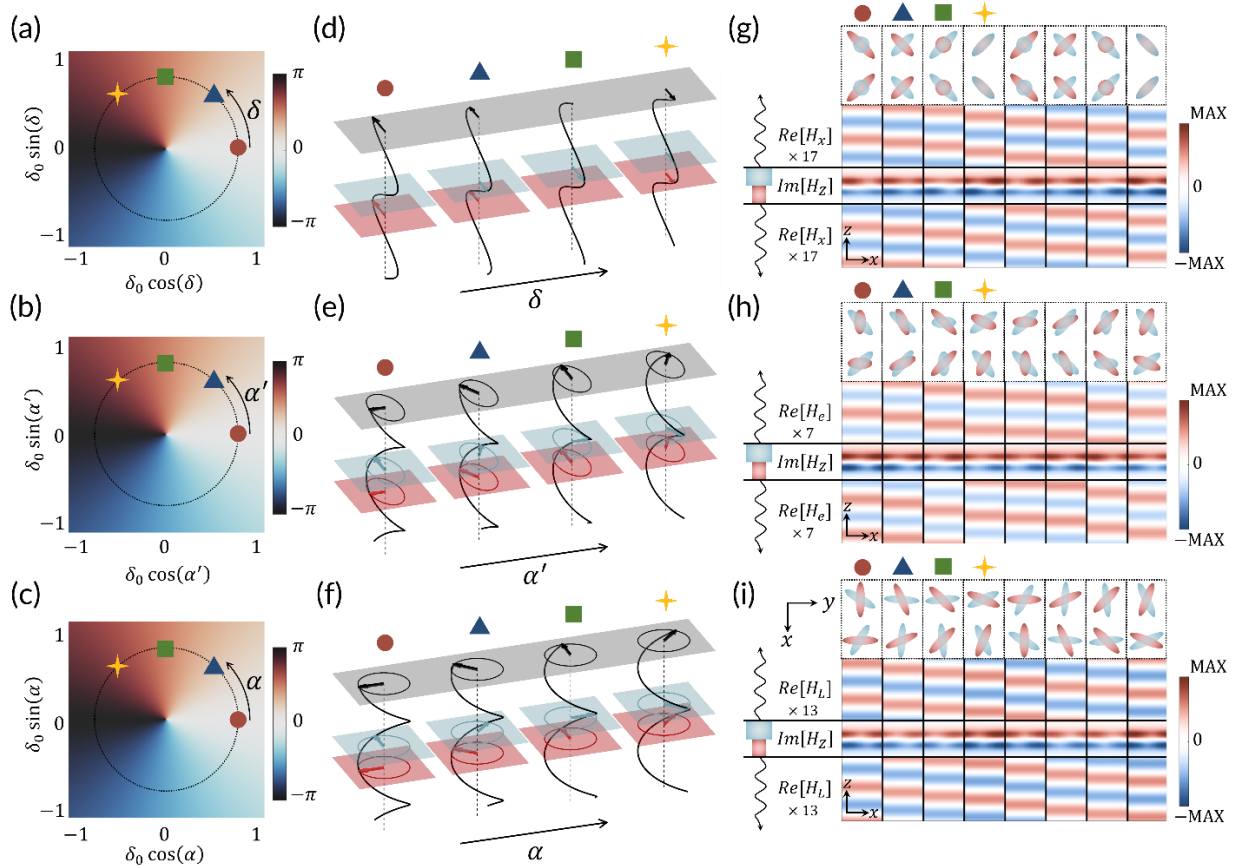

**Fig. 3**. **The decay phase: a polarization-agnostic geometric phase intrinsic to q-BICs.** TCMT predictions of the decay phase for (a) LP, (b) EP, and (c) CP light. (d-f) Geometric interpretation of the upward decaying wave of four select meta-units from (a-c), depicting the generalized geometric phase involving rotation in a multi-dimensional parameter space (with rotation parameters $\delta$ for LP, $\alpha'$ for EP, and $\alpha$ for CP). (g-i) Full-wave simulations showing the decay of the q-BICs (populated in the near field by a magnetic dipole source) for eight meta-units spanning $2\pi$ decay phase for the three polarizations in (a-c). For each scenario, the imaginary part of the q-BIC field $H_z$ is shown within the device (between black lines), and the real parts of the target polarization are shown decaying upwards and downwards. Colored markers (circle, triangle, square, star) track the four example meta-units across the left, middle, and right panels.

To enable control over general EP states, we consider how the q-BIC decays when directly populated by a dipole source placed in its near field, allowing direct study of the coupling coefficients $|d_{up}\rangle$ and $|d_{down}\rangle$ with full-wave simulations. We explore the decay with meta-units designed for three polarizations: $y$ polarization [Fig. 3(a)], an EP state with Jones vector $[1,-2i]/\sqrt{5}$ [Fig. 3(b)], and a CP state [Fig. 3(c)]. From the TCMT results, we identify the geometric parameter that controls the decay phase, $\Phi_d$. In the general EP case, this involves a change of variables from $(\delta,\alpha_1,\alpha_2)$ to $(\alpha',\beta',\gamma')$, such that $\alpha'$ controls the phase while $(\beta',\gamma')$ are kept constant to preserve $\psi$ and $\chi$. We find that $\Phi_d=\delta$ for LP, $\Phi_d=\alpha'$ for EP, and $\Phi_d=\alpha$ for CP, where

$$\tan(\alpha')=\frac{B-2\cos(2\delta)}{2\sin(2\delta)\cos(2\alpha_1-2\alpha_2)}$$
$$B=\sqrt{3+\cos(4\alpha_1-4\alpha_2)+2\cos(4\delta)\sin(2\alpha_1-2\alpha_2)}. \quad (3)$$
$$\alpha=\frac{\alpha_1+\alpha_2}{2}$$

As shown in Figs. 3(a-c), perturbing an a-BIC into a q-BIC in the appropriate parameter space reveals a topological feature of order 1, irrespective of the chosen polarization.

The introduced polarization-agnostic geometric phase has a simple geometric interpretation with respect to the direct and indirect scattering, as shown in Fig. 3(d-f). For CP light [Fig. 3(f)], the magnitudes of direct and indirect scattering are constant, while a rotation of the constituent LPs tunes the geometric phase (i.e., by varying $\alpha_1,\alpha_2$). For EP light [Fig. 3(e)], the magnitudes of the LPs are scaled while rotating the LP basis to maintain the correct ellipticity; this implies a composite rotation involving all three angles, $(\delta,\alpha_1,\alpha_2)$, as seen in $\alpha'$ in Eqn. (3) (see Supplementary Materials for $\beta'$ and $\gamma'$). For LP light [Fig. 3(d)], this scaling procedure is taken to the extreme, collapsing the ellipse onto a line because $(\alpha_1,\alpha_2)$ are constant. We confirm this picture with full-wave simulations, shown in Fig. 3(g-i). An arbitrary decay phase varying across $2\pi$ is indeed encoded by each library of meta-units, confirming that our proposal introduces a polarization-agnostic geometric phase.

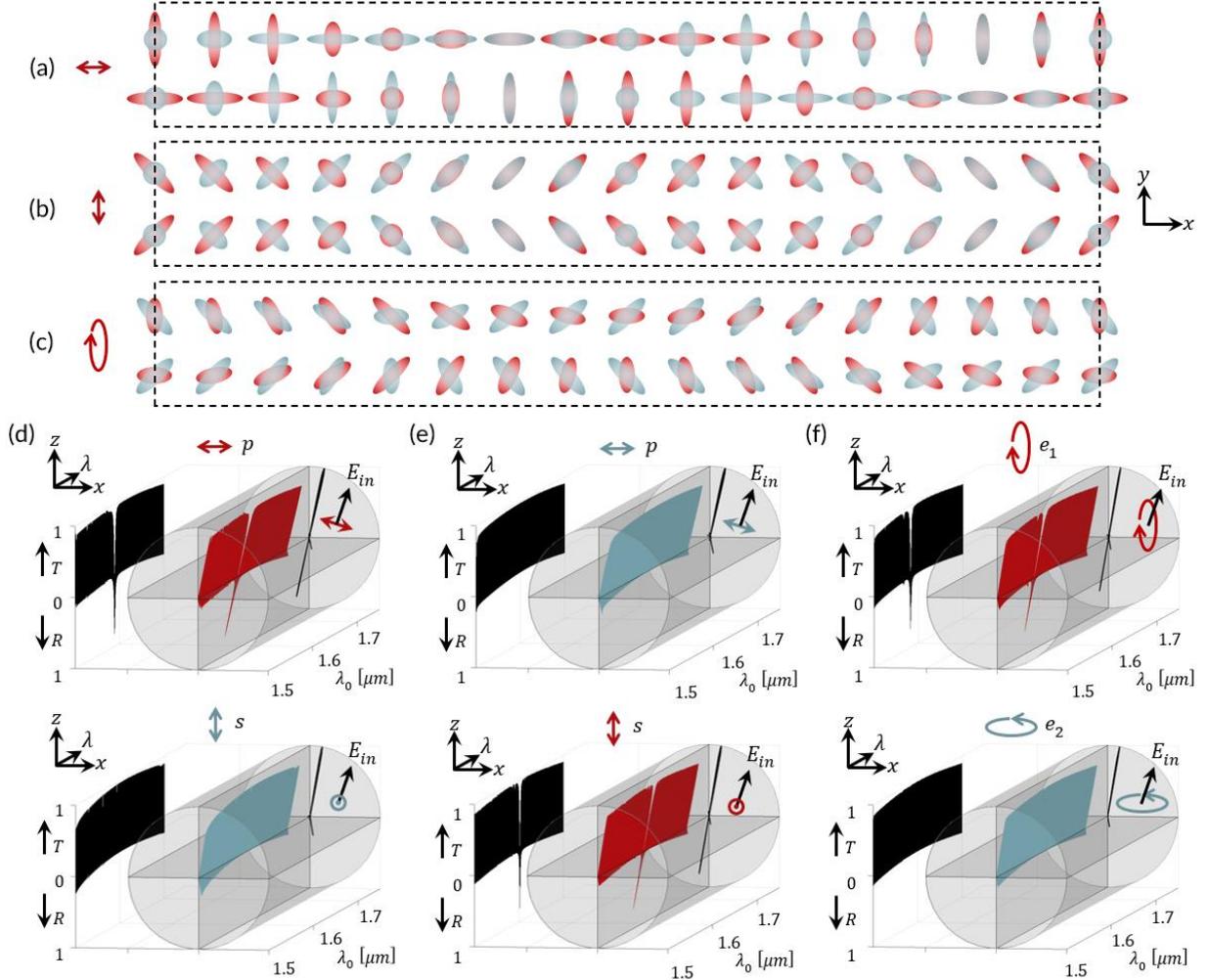

**Fig. 4**. **Decay phase gradients for arbitrary polarization.** Top-view of the super-period of nonlocal metasurfaces with geometric phase gradients, selective to (a) $p$ polarization, (b) $s$ polarization, and (c) and an elliptical polarization with Jones vector $[1,-2i]/\sqrt{5}$. (d) Far-field projections from full-wave simulations of the device in (a) when $p$ (top) and $s$ (bottom) polarizations are incident at an angle $10°$. (e) Same as (d) but the for the device in (b). (e) Same as (d) but for the device in (c) for the designed elliptical state (top) and orthogonal state (bottom).

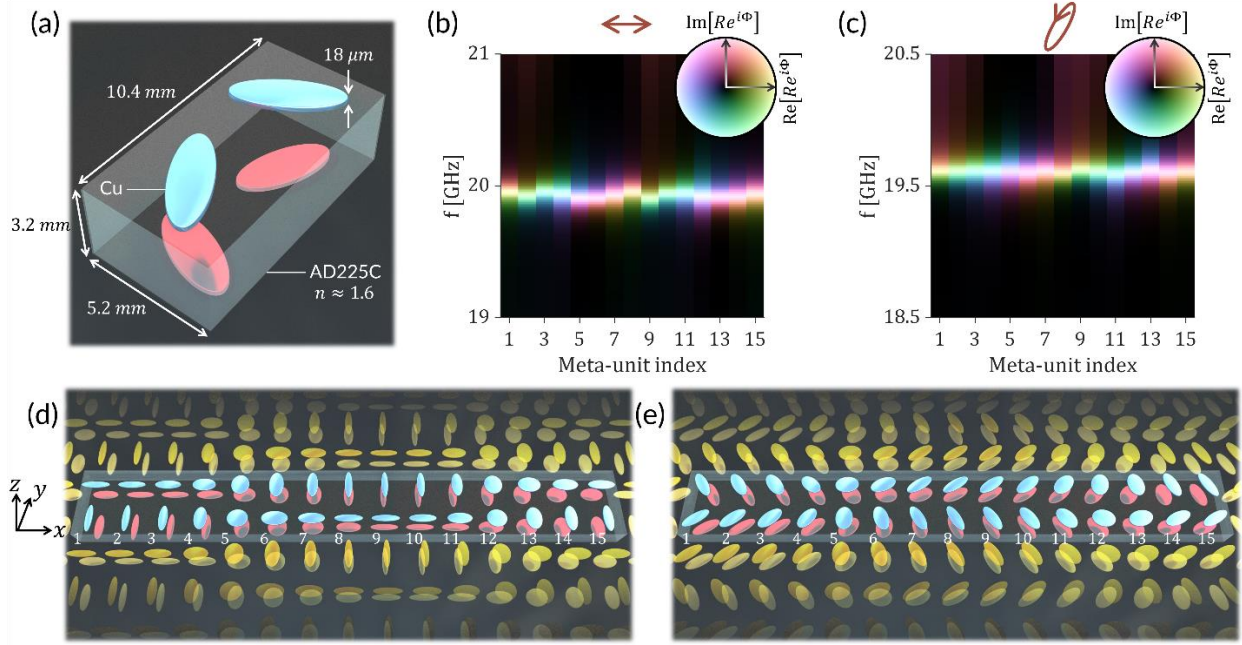

**Fig. 5. Nonlocal metasurfaces for radiofrequency operation.** (a) Meta-unit geometry based on copper ellipses on the top and bottom of an AD225C substrate. Meta-unit libraries for (b) $x$ polarization and (c) EP with $\psi = 60°, \chi = -15°$. In the colormaps, the value gives the reflectance, $R$, while the hue gives the reflected phase, $\Phi$. Nonlocal phase gradient metasurfaces for (d) $x$ polarization and (e) EP.

## Decay phase gradients

We are now ready to introduce a perturbation gradient across the aperture to encode custom profiles of the decay phase. Figure 4(a-c) show three linear phase gradient devices having superperiods of 16 meta-units, targeting three example polarizations. Figure 4(d) shows full-wave simulations of the transmitted and reflected far-field responses to *p*-polarized light at an incident angle of $10°$, exhibiting specular transmission punctuated by resonant retroreflection. Meanwhile, *s*-polarized light negligibly engages the resonance [Fig. 4(d)]. Conversely, based on the library in Fig. 3(g), Fig. 4(e) shows selective retroreflection of *s*-polarized light; similarly, based on the library in Fig. 3(h), Fig. 4(g-i) shows selective retroreflection for an EP state. CP retroreflection was demonstrated in Ref. [38] based on Ref. [37]. Detailed device parameters are reported in the Supplementary Materials.

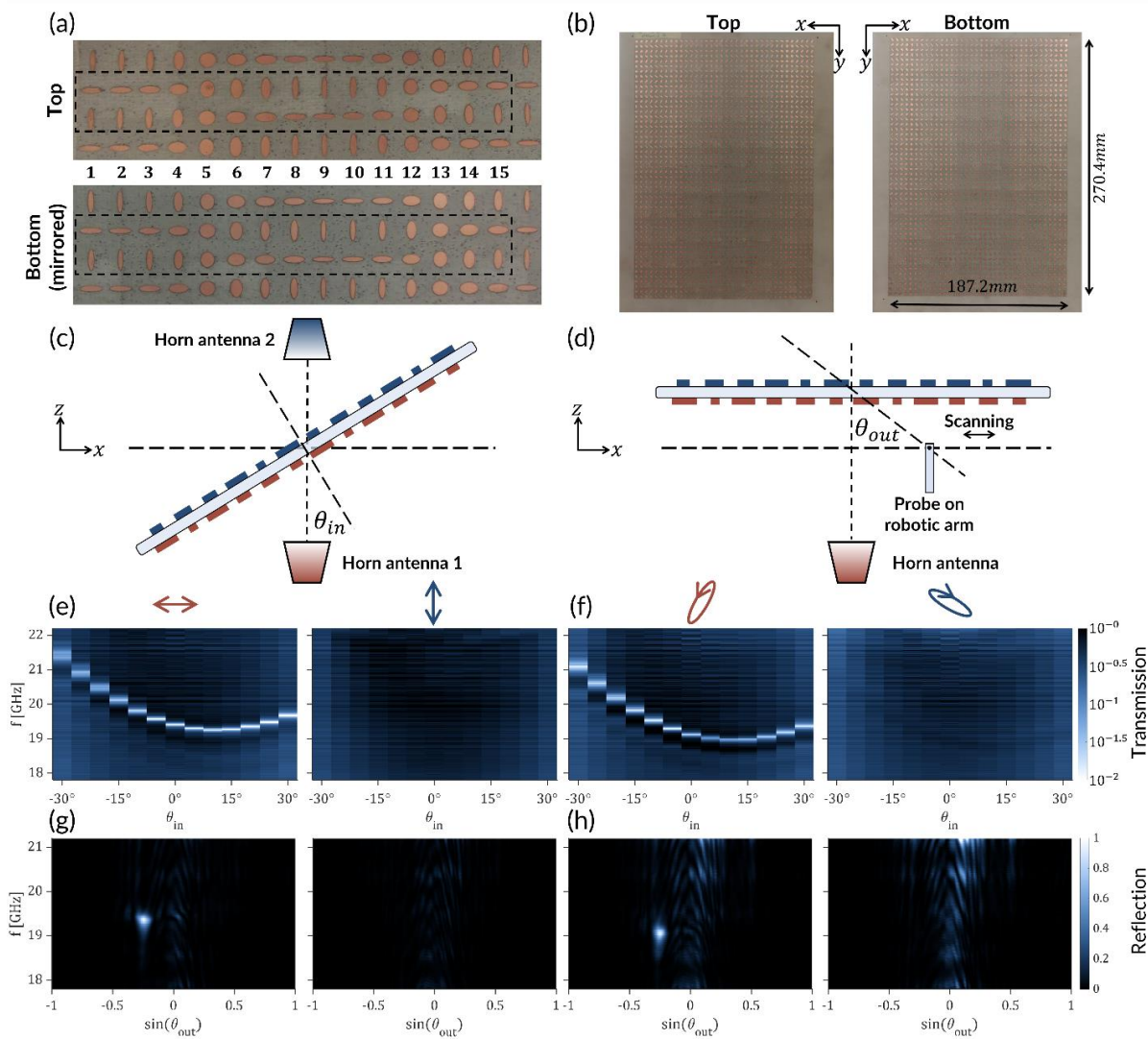

**Fig. 6**. **Experimental demonstration of polarization-agnostic nonlocal phase gradients.** (a) Images of the top and bottom layers of the fabricated LP devices, showing a superperiod of the device (dashed box). (b) Images of the top and bottom layers of the fabricated EP device, showing the entire array. The fabricated devices have 36 unit cells in the phase gradient direction and 26 unit cells in the orthogonal direction, (a total aperture of $187.2mm \times 270.4mm$ ). Experimental setups for (c) transmission and (d) reflection measurements. (e) Measured transmission for the LP device as a function of incident angle, $\theta_{in}$, for the designed polarization (left) and the orthogonal polarization (right). (f) Same as (e) but for the EP device. (g) Normalized reflection resolved as a function of outgoing angle,

$\theta_{out}$, for the designed polarization (left) and the orthogonal polarization (right) for the LP device when $\theta_{in}=0°$. (h) Same as (g) but for the EP device.

We experimentally demonstrate the introduced polarization-agnostic geometric phase in diffractive nonlocal metasurfaces operating near $20 GHz$, using a meta-unit made of copper and a dielectric substrate (AD225C) [Fig. 5(a)]. Based on this platform, we constructed a meta-unit library for $x$ polarization [Fig. 5(b)] and for an EP state [Fig. 5(c)] in which the Q-factor and resonant frequency are nearly constant while the reflection phase varies across $4\pi$. We may then array the 15 meta-units into phase gradients for LP [Fig. 5(d)] and EP [Fig. 5(e)] light.

We then fabricated phase gradients with superperiods of 15 meta-units to validate our theoretical results [Fig. 6(a,b)], measuring the transmission as a function incident angle for each polarization [Fig. 6(c)] and the reflection as a function of outgoing angle when excited from near-normal incidence [Fig. 6(d)] (details are in the Method section). Figures 6(e,f) show the measured polarization-resolved transmission varying the incident angle $\theta_{in}$, demonstrating momentum-shifted, polarization-selective bands. While typically a parabolic band is symmetric and centered about $\theta_{in}=0°$, here the q-BIC band structure is translated by a factor $k_g = -\frac{d\Phi_d}{dx}$. This shift is a hallmark of a nonlocal phase gradient [34],[38]. Remarkably, despite the finite size of the device and nontrivial decay phase profile, the spectrum dips to a residual transmission of only $\sim 1\%$ – a nearly complete Fano resonance at a custom angle. Meanwhile, the polarization selectivity is maintained across both devices, yielding no significant feature for the orthogonal polarizations. Finally, Figs. 5(g,h) show polarization-resolved reflection measurements as a function of reflected angle $\theta_{out}$ for a normally incident excitation. Here, we observe the a hallmark of nonlocal phase gradients: polarization-selective anomalous reflection exclusively at resonance [37]. While small nonresonant features are observed centered about the specular direction, at the resonant frequency, the energy is redirected by a momentum kick of $2k_g$. Together, these results confirm that the decay phase is imparted both upon coupling in, and again upon coupling out – these nonlocal metasurface completely engineer the spatial decay properties of the q-BIC supported by the slab.

## Outlook and conclusions

In this work, we introduced a polarization-agnostic geometric phase – the decay phase – enabling complete rational control over the four DoF defining a Fano resonance at normal incidence: the lifetime, phase and polarization state $(Q, \Phi, \psi, \chi)$. The results extend the capabilities of nonlocal metasurfaces, and at the same time add insights into their general operation. In particular, our analysis highlights that the phase at resonance for the reflected light is *twice* the decay phase ($\Phi = 2\Phi_d$): the reflected phase at resonance inherits two factors of decay phase—one from coupling into the q-BIC [the bra in Eqn. (1), representing time-reversed decay], and a second one upon coupling out (the ket, representing decay). So, while the topological order of the reflectance singularity in Fig. 2(b) is two (with phase accruing of $4\pi$), the order of the decay phase in Fig. 3(a) is one (phase accruing $2\pi$). We have hence shown that the decay phase of the q-BIC is the underlying mechanism responsible for the unique response of these systems. This view bolsters our understanding, originally shown in [34], of why a nonlocal phase gradient anomalously reflects to the $m = \pm 2$ diffraction orders (varying from $0$ to $4\pi$), while the near field of the q-BIC contains no phase discontinuity (varying from $0$ to $2\pi$). In contrast, conventional phase gradients in metasurfaces (e.g., based on varying the diameter of dielectric pillars) couple light to $m = \pm 1$ ($0$ to $2\pi$), suggesting the presence of near-field phase discontinuities ($0$ to $\pi$) that may lower the efficiency. Here, the continuity of the near field, along with the required lateral redistribution of energy [29], enables anomalous redirection of energy with arbitrarily high efficiency.

Notably, the observed responses are highly *selective* to the incoming frequency, wavefront and polarization — a hallmark of nonlocality [34]. Q-BICs are an ideal platform to realize *frequency selectivity*, due to their arbitrarily small and tailorable linewidths, combined with an independent broad non-resonant response. Here, we studied broadband transparency, which enables highly multifunctional meta-optics by cascading several devices [40]. In addition, our spatially varying geometric phase allows customizable *wavefront selectivity* to an incoming wave of choice [38]. Here, our results show that the underlying mechanism is in fact the decay phase, rather than the conventional PB phase. Finally, in this work we generalized these concepts to arbitrary EP states, adding *polarization selectivity*, and showing that the introduced decay phase is agnostic to polarization. Since it accumulates upon encircling a topological feature in a geometric parameter space, it is a clear manifestation of the geometric phase [19]. Based on these findings, we have been able to experimentally demonstrate for the first time a geometric phase for linear and elliptical states.

Our results suggest several paths forward for engineering open systems using symmetry-breaking concepts. For instance, in parallel to this work we anticipated a distinct manifestation of the polarization-agnostic geometric phase in single-layer grating couplers [49]. While these devices emit mirrored copies of the designed wavefront upwards and downwards, our results here suggest that two-layer grating couplers may command radiation to the two half-spaces with striking flexibility. Relatedly, since the decay symmetries to the two half-spaces determines the modal dynamics of two spectrally overlapping states [50], our work opens unexplored opportunities for multi-state interference in open, quasi-2D systems. Most generally, our results imply the generalization of wavefront selectivity observed in nonlocal metasurfaces [38] to vector-beam selectivity: external light engaging a Fano resonant state exclusively for specifically tailored amplitude, phase and polarization profiles. This complete platform for nonlocal metasurfaces is capable of unprecedented multifunctionality through cascading surfaces that target specific wavelengths, polarizations and wavefront profiles, ideally suited for augmented reality, secure communications, wireless and optical communications, next generation integrated photonics and antenna systems. Due to the broad applicability of symmetry-driven design, geometric phases, and topological phenomena, our findings may manifest in a wide range of physical systems, such as acoustics, elastics, and quantum materials.

## Methods

**Numerical simulations**

The simulations for the all-dielectric metasurfaces in Figs. 2-4 were carried out using the Finite Difference Time Domain method using the commercial software, Lumerical solutions.

Nondispersive refractive indices were used for all materials. The ring-down method was used to determine the Q-factors in Fig. 2(b), while the reflection phase was retrieved from a monitor behind a free-space source. The a-BIC profile in Fig. 2(c) was simulated using a dipole in the near field in the case with no perturbation. Similarly, the full-wave simulations in Fig. 3(g-i) were computed using a dipole source in the near field, and monitors recording the magnitude field at an $xz$ plane. The full-wave simulations in Fig. 4(d-f) were performed by exciting the device from the substrate at an angle of incidence $10°$ off the device normal, for each of the polarizations shown. The spatial distribution of the fields in the reflection and transmission regions were recorded and post-processed by projecting to the far fields for each wavelength.

The simulations for the metallic metasurfaces in Fig. 5 were carried out using the commercial finite element method solver, Ansys HFSS. The geometry of the meta unit in Fig. 5(a) is modeled assuming an infinite 2D array using periodic boundaries and Floquet ports under the condition of normal incidence. Calculated $S$-parameters were post-processed and transformed to the reflectance in the designed polarization states.

**Device fabrication**

Two nonlocal metasurfaces in Fig. 5 (d-e) were fabricated using a circuit board prototyping machine, LPKF ProtoLaser S4. The substrate material used is ADD255C (with relative permittivity $\varepsilon_\mathrm{r} = 2.55$ and loss tangent $\tan\delta = 0.0014$). The thicknesses of the dielectric and the copper cladding are 3.2 $mm$ and 18 $\mu m$, respectively. In the fabricated metasurfaces, the phase gradient is encoded in one direction while the other direction is periodic. 36 and 26 meta units are arrayed along the phase-gradient and the dimerized direction, respectively, giving the aperture area of 187.2 mm × 270.4 mm (the size of a meta-unit is 5.2 mm×10.4 mm). This aperture area is patterned approximately the center of the printed circuit board with the dimension of 229mm × 305mm. Images of the fabricated devices are supplied in Fig. 6 (a,b) and also in the Supplementary Materials, Figure S6.

**Device characterization**

The fabricated devices were characterized by two types of measurements: transmission and reflection measurements. Figure S7(a) shows an image of the setup for transmission measurements, while Figure S8(a) shows an image of the setup for reflection measurements. The momentum-shifted bands in Fig. 6 (e-f) were captured by measuring the transmissive *S*-parameters. The measurement setup consists of two standard horn antennas (Pasternack PE-9852/2F-15) facing each other and placed 1.8 meters apart. The nonlocal metasurfaces is placed at the center of the antennas on an azimuthal rotational stage. The rotational stage allows to change the incident angle in the phase gradient direction. The transmissive *S*-parameters were measured using a vector network analyzer (Rohde & Schwarz ZVA-40) for every combination of excitation polarization *i* and analyzing polarization *j*, where *i*, *j* form a linear basis. This cycle of the measurement was repeated by rotating the rotational stage to change the incident angle, $\theta_{\text{in}}$, from −40° to 40° with a 5° step. The angle of the normal incidence was calibrated by measuring a uniform nonlocal metasurface (a device with a circular polarization state fabricated in [45] was used), which has a nonshifted band in momentum space (i.e., the band edge exists at normal incidence, corresponding to $\theta_{in} = 0°$). For post-processing, the measured *S*-parameters are normalized by *S*-parameters captured in a setup where the nonlocal metasurface is simply removed from a metasurface holder.

Then, for the nonlocal metasurface with the elliptic polarization state [Fig. 5 (g)], the normalized *S*-parameters were transformed into the transmittance in the desired polarization basis.

The reflection measurement was performed approximately under the normal incidence condition. The nonlocal metasurface is illuminated by the standard horn antenna placed about 1.9 meters away from the metasurface. The reflected field from the metasurface was captured point-by-point in a line parallel to and 230 $mm$ away from the metasurface by a waveguide probe (Eravant SAP-42-R2) using a 3D robotic-arm position scanner (Fanuc LR Mate 200iC). The scanned horizontal line extends to 600 $mm$ along the phase-gradient direction with a scanning step of 5 $mm$ (121 points), and is located 45 $mm$ above the center of the feed horn, meaning we captured the reflected-field components with a slightly tilted incident angle, $\left[\theta_{in} = \mathrm{asin}\left(\frac{45\mathrm{mm}/2}{1.9\mathrm{m}}\right) \lesssim 1°\right]$ in the dimerized direction. As in the transmission measurement, this measurement cycle was repeated to obtain the data for every combination of linear polarizations $i$ and $j$. The measured position-dependent data are transformed into the momentum space by performing a discrete Fourier transform. A change of polarization basis was then applied to transform from a linear basis to the designed polarizations for the nonlocal metasurfaces with elliptical polarization state [Fig. 5 (i)]. The results were then normalized to the peak reflectance in each case.

**Acknowledgements**

This work was supported by AFOSR and the Simons Foundation.

**Author contributions**

A.C.O conceived the ideas. A.C.O designed and performed simulations for the infrared devices. Y.K. designed, simulated, and fabricated the RF devices. Y.K. and G.X. performed the experiments. Y.K. and A.C.O. analyzed the data. A.C.O, Y.K., and A.A. wrote the manuscript. A.A. supervised the research.

**Competing interests**

The authors declare no competing interests.

**Supplementary Materials for:**

# Demonstration of a Polarization-Agnostic Geometric Phase in Nonlocal Metasurfaces

Adam Overvig[1,*], Yoshiaki Kasahara[1,2,*], Gengyu Xu[1], and Andrea Alù[1,3,†]

[1]Photonics Initiative, Advanced Science Research Center at the Graduate Center of the City University of New York

[2]Department of Electrical and Computer Engineering, the University of Texas at Austin

[3]Physics Program, Graduate Center, City University of New York

* These authors contributed equally. †Corresponding author: aalu@gc.cuny.edu

## S1. Degrees of freedom

We consider the number of independent degrees of freedom (DoF) available in nonlocal metasurfaces excited at normal incidence whose broadband (background) response is near-unity transmission, punctuated by a narrowband Fano resonance that reflects with the Jones matrix with 8 DoF:

$$J_r = \begin{bmatrix} r_{xx} & r_{xy} \\ r_{yx} & r_{yy} \end{bmatrix}. \tag{1}$$

In reciprocal devices, we must have $J_r = J_r^T$ (enforcing $r_{xy} = r_{yx}$), leaving 6 DoF. In this work, we are interested in the DoF engaged by a single resonant state controlled by perturbations to a high symmetry lattice (e.g., no birefringence), in which case a single polarization state $|e_1\rangle$ engages the resonance, while the orthogonal state $|e_2\rangle$ does not. That is, $J_r|e_2\rangle = 0$, removing a further 2 DoF to yield 4 DoF. In this case, a simple decomposition of $J_r$ gives

$$J_r = |e^*_1\rangle\langle e_1|. \tag{2}$$

Here, $|e_1\rangle$ has 4 DoF, which we may view as a magnitude $\sqrt{A}$ (where $A^2$ is the peak reflectance), phase $\Phi$ and polarization state (with elliptical parameters $\psi, \chi$) giving a set $(A, \Phi, \psi, \chi)$. In the absence of loss, the reflectance at resonance must be unity in two-port systems, removing the amplitude as a DoF (i.e., $A=1$). However, when there is loss (quantified by a nonradiative scattering rate $\gamma_{nr}$), the peak reflectance varies as $A^2 = 2\gamma_r^2/(\gamma_{nr}+\gamma_r)^2$, where $\gamma_r = \omega_0/Q$ is the radiative scattering rate for a resonant frequency $\omega_0$ and hence $Q$ can be seen as controlling the amplitude for fixed $\gamma_{nr}$. Since the Q-factor is an important DoF of the system even in the lossless cases $\gamma_{nr} = 0$, it is more natural to consider it as a DoF in the place of amplitude. In this way, we arrive at the four DoF of interest: $(Q, \Phi, \psi, \chi)$. In this work, we show a system with rational control over all four DoF simultaneously, completing the program of commanding the decay of a single q-BIC begun in Refs. [37,38] of the main text. A system with simultaneous control of $\gamma_{nr}$ would readily extend this one step further to control both Q-factor and amplitude; here, we consider the lossless case $\gamma_{nr} = 0$.

## S2. Temporal coupled mode theory (TCMT)

We begin with the dynamical equations for a q-BIC with complex modal amplitude $a(t)$ excited by an incoming wave $|s_+\rangle$, and aim to determine the outgoing waves $|s_-\rangle$. Here, $|a(t)|^2$ is normalized to be the energy stored in the q-BIC per unit area at time $t$ while $\langle s_+|s_+\rangle$ ($\langle s_-|s_-\rangle$) is normalized to be the incident (outgoing) intensity. The dynamical equations read

$$-i\omega a = -(i\omega_r + \gamma_r + \gamma_i)a + \langle d^*|s_+\rangle$$
$$|s_-\rangle = C|s_+\rangle + |d\rangle a$$
(3)

where the coupling vector $|d\rangle$ contains the coupling coefficients to and from free-space in some appropriate basis, $\omega_r$ is the (resonant) modal frequency of the q-BIC, $\gamma_r$ is the radiative scattering rate, and $\gamma_i$ is the nonradiative scattering rate (which we will take to be 0 throughout). The scattering rates are interchangable with their corresponding lifetimes, e.g. $\tau_r = 1/\gamma_r$ is the radiative lifetime of the mode.

In this work, we are interested in the local scattering matrix of the form:

$$C = e^{i\Phi_c}\begin{bmatrix} r_0 & 0 & -it_0 & 0 \\ 0 & r_0 & 0 & -it_0 \\ -it_0 & 0 & r_0 & 0 \\ 0 & -it_0 & 0 & r_0 \end{bmatrix},$$
(4)

where $r_0$ and $t_0$ are Fresnel coefficients, and the reference phase $\Phi_c$ is chosen such that $r_0, t_0$ are real-valued. In the main text, we assume $r_0 = 0$ and $t_0 = 1$, but we keep them general for now. The electric field basis for this matrix is:

$$E = \begin{bmatrix} E_{1,x} \\ E_{1,y} \\ E_{2,x} \\ E_{2,y} \end{bmatrix},$$
(5)

where $E_{i,p}$ is the electric field on side $i$ with polarization $p$. TCMT aims to determine the coefficients of the coupling vector

$$|d\rangle = \begin{bmatrix} d_1 \\ d_2 \\ d_3 \\ d_4 \end{bmatrix} e^{i\Phi_c/2}, \tag{6}$$

where we include the factor phase factor $e^{i\Phi_c/2}$ for convenience because it yields

$$S = e^{i\Phi_c} \left\{ \begin{bmatrix} r_0 & 0 & -it_0 & 0 \\ 0 & r_0 & 0 & -it_0 \\ -it_0 & 0 & r_0 & 0 \\ 0 & -it_0 & 0 & r_0 \end{bmatrix} + \frac{1}{i(\omega-\omega_0)-\gamma_r} \begin{bmatrix} d_1 d_1 & d_1 d_2 & d_1 d_3 & d_1 d_4 \\ d_2 d_1 & d_2 d_2 & d_2 d_3 & d_2 d_4 \\ d_3 d_1 & d_3 d_2 & d_3 d_3 & d_3 d_4 \\ d_4 d_1 & d_4 d_2 & d_4 d_3 & d_4 d_4 \end{bmatrix} \right\}. \tag{7}$$

In this form we may drop the reference phase $\Phi_c$.

Next, we need to determine the phenomenological form of $|d\rangle$ relevant to the geometry in Fig. S1 and then apply the constraints of conservation of energy, reciprocity, and time-reversal symmetry to determine the coefficients in $|d\rangle$. We begin with writing down the correct form for $|d\rangle$ based on the selection rules. The perturbation of interest [Fig. S1(a)] is characterized by four parameters: the perturbation at the bottom and top interfaces have strengths $\delta_1$, and $\delta_2$, respectively, while the orientation angles are $\alpha_1$ and $\alpha_2$ for the bottom and top interfaces, respectively. As depicted in Fig. S1(d), the phenomenological translation of this device is as follows: the perturbation at the bottom interface scatters light to polarization $\phi$ with a strength proportional to the magnitude of the perturbation, $\delta_1$; the perturbation at the top interface scatters light to polarization $\theta$ with a strength proportional to the magnitude of that perturbation, $\delta_2$. The polarization angles follow

$$\phi \approx 2\alpha_1$$
$$\theta \approx 2\alpha_2, \tag{8}$$

and, as motivated in the main text, we use the parameterization

$$\begin{aligned}\delta_1 &= \delta_0 \cos(\delta)\\ \delta_2 &= \delta_0 \sin(\delta)\end{aligned}. \tag{9}$$

Here, $\delta_0$ is the collective perturbation strength while $\delta$ is an angular parameter determining the relative perturbation strength contributed by each interface. The Q-factor of the q-BIC is then controlled as $Q \propto 1/\delta_0^2$, by construction.

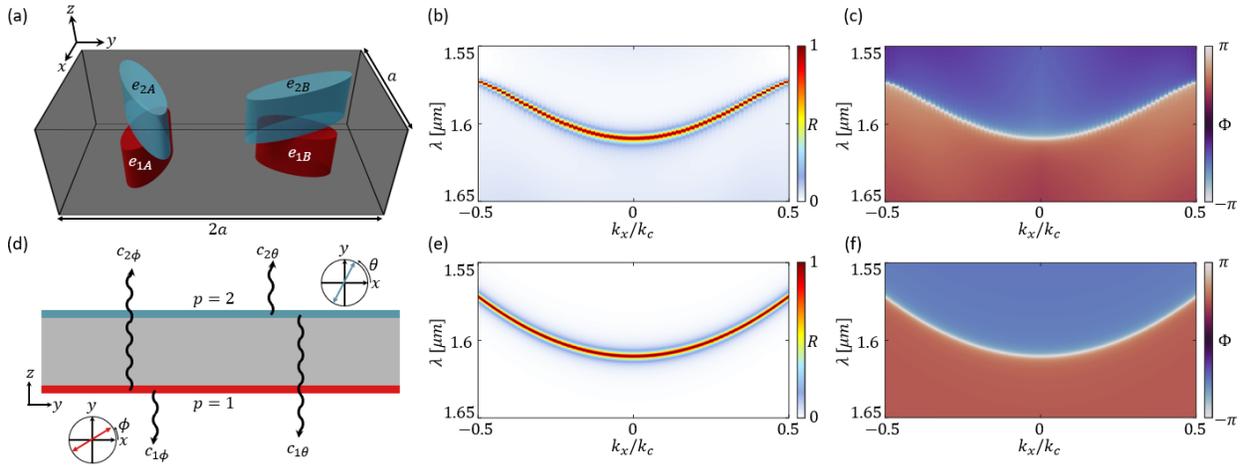

**Figure S1. Comparison of a nonlocal metasurface implementation studied by full-wave simulations (a-c) and modeled by TCMT (d-f).** (a) Unit cell of a nonlocal metasurface constructed from four elliptical inclusions. (b,c) Reflection and reflected phase for the case in (a) where the ellipses are aligned in the y direction ($\alpha_1 = \alpha_2 = 0$). (d) Schematic of the TCMT model of the device in (a). (e,f) Reflection and reflected phase from a TCMT fit to match the device in (a-c) near $k_x = 0$.

Next, we write the elements of $|d\rangle$ based on the appropriate sum of scattering events from each interface, broken into *x* and *y* components based on the polarization angles $\phi$ and $\theta$

$$d_1 = c_{1\phi} \cos(\phi) + c_{1\theta} \cos(\theta)$$
$$d_2 = c_{1\phi} \sin(\phi) + c_{1\theta} \sin(\theta)$$
$$d_3 = c_{2\phi} \cos(\phi) + c_{2\theta} \cos(\theta)$$
$$d_4 = c_{2\phi} \sin(\phi) + c_{2\theta} \sin(\theta)$$

(10)

Here, the coefficients $c_{ip}$ are the complex scattering coefficients to port $i$ from the scatterer producing a polarization $p$. However, due to the symmetry of the device (rigorous in the case of vanishing perturbation strength, approximate in the case of the presence of a perturbation), these coefficients are not independent. Instead, we write the indirect scattering components $c_{2\phi}$ and $c_{1\theta}$ in terms of the direct scattering components $c_{1\phi}$ and $c_{2\theta}$:

$$c_{1\phi} = d_0 \cos(\delta) e^{i\Phi}$$
$$c_{2\theta} = \pm d_0 \sin(\delta) e^{i\Phi}$$
$$c_{2\phi} = c_{1\phi} g e^{i\Delta\Phi}$$
$$c_{1\theta} = c_{2\theta} g e^{i\Delta\Phi}$$

(11)

Here, the plus (minus) signs pertain to symmetric (anti-symmetric) modes. We will proceed assuming a symmetric mode, noting that an anti-symmetric mode may also be obtained by the replacement $\delta \to -\delta$. In the general case, the direct scattering coefficients need not be equal in magnitude (the relative magnitude is parameterized by $\delta$). However, by the symmetry of the unperturbed system (and thereby, the mode profile), the phase $\Phi$ will be the same. The indirect scattering coefficients here are written using a relative amplitude $g$ and phase factor $\Delta\Phi$. The phase factor inherits the symmetry of the unperturbed structure, and is $\Delta\Phi = -\pi/2$, corresponding to the phase relation of the reflected (direct) and transmitted (indirect) light in Eqn. (4).

The unknowns in (11) are also subject to the physical constraints of energy conservation,

$$\langle d|d\rangle = 2\gamma_r \tag{12}$$

and time-reversal symmetry,

$$C|d^*\rangle = -|d\rangle. \tag{13}$$

From Eqn. (12) we must have

$$|d_1|^2 + |d_2|^2 + |d_3|^2 + |d_4|^2 = 2\gamma_r \tag{14}$$

which leads to

$$|d_0|^2 = \frac{2\gamma_r}{1+g^2}. \tag{15}$$

And from Eqn. (13), we must have

$$r_0 d_1^* - it_0 d_3^* = -d_1. \tag{16}$$

which leads to

$$\frac{r_0\left(\cos(\phi)\cos(\delta)+i\cos(\theta)\sin(\delta)g\right)-it_0\left(i\cos(\phi)\cos(\delta)g+\cos(\theta)\sin(\delta)\right)}{\cos(\phi)\cos(\delta)-i\cos(\theta)\sin(\delta)g} = -e^{2i\Phi}. \tag{17}$$

Denoting the left-hand side as $h_0$, we see that Eqn. (17) demands that $|h_0|=1$, which, after some simplification, enforces

$$g = \frac{1-|r_0|}{\sqrt{1-r_0^2}}. \tag{18}$$

This is a crucial result, showing that the contribution of indirect scattering is not independent of the background scattering. In particular, when the background reflectance is unity, $r_0 = 1$, we have no contribution from indirect scattering, $g = 0$. However, when the background transmission is unity, $t_0 = 1$, the contribution from indirect scattering is equal in magnitude to that of the direct scattering, $g = 1$. To be clearer: in the latter case, the direct and indirect scattering have equal amplitude for *the same magnitude of perturbation*, yet have a relative phase between them of $90°$. For this reason, we are interested in device operating near the transmission peak of a Fabry-Perot background, where $r_0 = 0$, yielding

$$\begin{aligned} g &= 1 \\ h_0 &= 1 \\ e^{i\Phi} &= i \\ |d_0| &= \sqrt{\gamma_r} \end{aligned} \tag{19}$$

Finally, we arrive at the form

$$\begin{aligned} d_1 &= i\sqrt{\gamma_r}\left[\cos(\phi)\cos(\delta) - i\cos(\theta)\sin(\delta)\right] \\ d_2 &= i\sqrt{\gamma_r}\left[\sin(\phi)\cos(\delta) - i\sin(\theta)\sin(\delta)\right] \\ d_3 &= i\sqrt{\gamma_r}\left[-i\cos(\phi)\cos(\delta) + \cos(\theta)\sin(\delta)\right] \\ d_4 &= i\sqrt{\gamma_r}\left[-i\sin(\phi)\cos(\delta) + \sin(\theta)\sin(\delta)\right] \end{aligned} \tag{20}$$

This result may be separated into the upward and downward coefficients

$$d_{up} = \begin{bmatrix} d_1 \\ d_2 \end{bmatrix} = i\sqrt{\gamma_r}\left\{\cos(\delta)\begin{bmatrix} \cos(\phi) \\ \sin(\phi) \end{bmatrix} - i\sin(\delta)\begin{bmatrix} \cos(\theta) \\ \sin(\theta) \end{bmatrix}\right\}$$

$$d_{down} = \begin{bmatrix} d_3 \\ d_4 \end{bmatrix} = i\sqrt{\gamma_r}\left\{-i\cos(\delta)\begin{bmatrix} \cos(\phi) \\ \sin(\phi) \end{bmatrix} + \sin(\delta)\begin{bmatrix} \cos(\theta) \\ \sin(\theta) \end{bmatrix}\right\}$$
(21)

This form clarifies that, relative to some phase, the real part of the upward scattered state is given by the linear polarization state $\phi$ with strength $\cos(\delta)$ while the (negative) imaginary part is given by the linear polarization state $\theta$ with strength $\sin(\delta)$.

With the elements of $|d\rangle$ known, the overall scattering is uniquely determined by the parameters $r, \omega_0, \tau_r, \phi, \theta$:

$$S = e^{i\Phi_c}\left[C + \frac{1}{1-i\Omega\tau_r}D\right],$$
(22)

where for our case we have

$$C = -i\begin{bmatrix} 0 & 0 & 1 & 0 \\ 0 & 0 & 0 & 1 \\ 1 & 0 & 0 & 0 \\ 0 & 1 & 0 & 0 \end{bmatrix},$$

$$D = \begin{bmatrix} d_1 \\ d_2 \\ d_3 \\ d_4 \end{bmatrix}\begin{bmatrix} d_1 & d_2 & d_3 & d_4 \end{bmatrix},$$

$$\begin{bmatrix} d_1 \\ d_2 \\ d_3 \\ d_4 \end{bmatrix} = \cos(\delta)\begin{bmatrix} \cos(\phi) \\ \sin(\phi) \\ -i\cos(\phi) \\ -i\sin(\phi) \end{bmatrix} + \sin(\delta)\begin{bmatrix} -i\cos(\theta) \\ -i\sin(\theta) \\ \cos(\theta) \\ \sin(\theta) \end{bmatrix},$$
(23)

$$\Omega = \omega - \omega_r,$$

$$\Phi_c = \pi,$$

Here, we note that $\Omega$ (the difference in incident and resonant frequencies) may be adjusted as

$$\Omega \to \Omega = \omega - (\omega_0 - i\gamma_i), \tag{24}$$

in order to account for the nonradiative loss rate $\gamma_i$. The resonant frequency may also be approximated as a function of in-plane momentum $k$ near normal incidence ($k=0$) as

$$\omega_r(k) \approx \omega_0 + \frac{b}{2}k^2, \tag{25}$$

where $\omega_0$ is the resonant frequency at normal incidence (the band-edge frequency) and $b$ is a Taylor expansion coefficient. Figure S1(b,c) show the reflection and phase calculated by FDTD for a device with $\alpha_1 = \alpha_2 = 0$, $H = 500nm$, and $a = 400nm$. Figure S1(e,f) shows the reflection using the TCMT model fit to the response of this device.

Finally, we parameterize $\phi, \theta, \delta, \gamma_r$ in the TCMT with the selection rules in terms of $\delta_0, \delta, \alpha_1, \alpha_2$ for the magnetic mode of interest:

$$\begin{aligned} \phi &\approx 2\alpha_1 \\ \theta &\approx 2\alpha_1 \\ \delta_{TCMT} &\approx \delta ' \\ \gamma_r &\propto \delta_0^2 \end{aligned} \tag{26}$$

where we have clarified that the $\delta$ used in the TCMT is in principle distinct from the geometric $\delta$, but in practice are approximately the same. These substitutions lead us to the form reported in the main text:

$$|d_{up}\rangle \propto i\delta_0 \left\{ \cos(\delta) \begin{bmatrix} \cos(2\alpha_1) \\ \sin(2\alpha_1) \end{bmatrix} - i\sin(\delta) \begin{bmatrix} \cos(2\alpha_2) \\ \sin(2\alpha_2) \end{bmatrix} \right\}$$
$$|d_{down}\rangle \propto i\delta_0 \left\{ -i\cos(\delta) \begin{bmatrix} \cos(2\alpha_1) \\ \sin(2\alpha_1) \end{bmatrix} + \sin(\delta) \begin{bmatrix} \cos(2\alpha_2) \\ \sin(2\alpha_2) \end{bmatrix} \right\}$$
(27)

from which we may easily compute the Jones matrix. For instance, the Jones matrix for light incident from side 2 is

$$J_r = \begin{bmatrix} r_{xx} & r_{xy} \\ r_{xy} & r_{yy} \end{bmatrix}$$
$$r_{xx} = \left[ \cos(\delta)\cos(2\alpha_1) - i\sin(\delta)\cos(2\alpha_2) \right]^2$$
$$r_{xy} = \left[ \cos(\delta)\cos(2\alpha_1) - i\sin(\delta)\cos(2\alpha_2) \right]\left[ \cos(\delta)\sin(2\alpha_1) - i\sin(\delta)\sin(2\alpha_2) \right]$$
$$r_{yy} = \left[ \cos(\delta)\sin(2\alpha_1) - i\sin(\delta)\sin(2\alpha_2) \right]^2$$
(28)

**S3. Mapping $(\delta_0, \delta, \alpha_1, \alpha_2)$ to and from $(Q, \Phi, \psi, \chi)$**

While there are many valid ways to parameterize the four degrees of freedom for light (amplitude, phase, polarization), here we describe the Jones vector using $(A, \Phi, \psi, \chi)$:

$$E = \begin{bmatrix} E_x \\ E_y \end{bmatrix} = Ae^{i\Phi} \begin{bmatrix} \cos(\psi) & -\sin(\psi) \\ \sin(\psi) & \cos(\psi) \end{bmatrix} \begin{bmatrix} \cos(\chi) \\ i\sin(\chi) \end{bmatrix},$$
(29)

where we may invert the Stokes vector

$$\begin{bmatrix} I \\ Q \\ U \\ V \end{bmatrix} = A^2 \begin{bmatrix} 1 \\ \cos(2\psi)\cos(2\chi) \\ \sin(2\psi)\cos(2\chi) \\ \sin(2\chi) \end{bmatrix} = \begin{bmatrix} |E_x|^2 + |E_y|^2 \\ |E_x|^2 - |E_y|^2 \\ 2\mathrm{Re}(E_x E_y^*) \\ -2\mathrm{Im}(E_x E_y^*) \end{bmatrix}$$
(30)

into

$$2\psi = \tan^{-1}\left(\frac{U}{Q}\right)$$

$$2\chi = \tan^{-1}\left(\frac{V}{\sqrt{Q^2 + U^2}}\right) \tag{31}$$

Since we aim to parameterize the *leakage* of a state, i.e., its coupling coefficients $|d\rangle$, we write $|d\rangle$ based on Eqn. (29)

$$|d\rangle = \sqrt{\gamma}e^{i\Phi}\begin{bmatrix} \cos(\psi) & -\sin(\psi) \\ \sin(\psi) & \cos(\psi) \end{bmatrix}\begin{bmatrix} \cos(\chi) \\ i\sin(\chi) \end{bmatrix}. \tag{32}$$

Where we have replaced $A$ with a scattering rate $\sqrt{\gamma}$, related to the Q-factor $Q = \omega_0/\gamma$. Notably, we could also choose to the parameterize light using the amplitudes ($A_{re}, A_{im}$) and angles ($\phi_{re}, \phi_{im}$) of the real and imaginary parts of the wave:

$$E = A_{re}\begin{bmatrix} \cos(\phi_{re}) \\ \sin(\phi_{re}) \end{bmatrix} + iA_{im}\begin{bmatrix} \cos(\phi_{im}) \\ \sin(\phi_{im}) \end{bmatrix} \tag{33}$$

or for the coupling coefficients,

$$|d\rangle = \sqrt{\gamma_{re}}\begin{bmatrix} \cos(\phi_{re}) \\ \sin(\phi_{re}) \end{bmatrix} + i\sqrt{\gamma_{im}}\begin{bmatrix} \cos(\phi_{im}) \\ \sin(\phi_{im}) \end{bmatrix}. \tag{34}$$

Inspection of Eqn. (27) shows that this closely matches the form predicted by TCMT, and equating for instance the downward component of Eqn. (27) straightforwardly prescribes

$$\begin{aligned}
\delta_1 &= \delta_0 \cos(\delta) = \sqrt{\gamma_{re}} \\
\delta_2 &= \delta_0 \sin(\delta) = \sqrt{\gamma_{im}} \\
\alpha_1 &= \phi_{re}/2 \\
\alpha_2 &= \phi_{im}/2
\end{aligned} \tag{35}$$

That is, the bottom interface controls the real part (scattering rate and orientation angle) and the top interface controls the imaginary part (scattering rate and orientation angle).

We may also prescribe the required $\delta_0, \delta, \alpha_1, \alpha_2$ based on the desired $(Q, \Phi, \psi, \chi)$ through

$$|d_{down}\rangle = \sqrt{C_0}\delta_0 \left\{ \cos(\delta)\begin{bmatrix}\cos(2\alpha_1)\\ \sin(2\alpha_1)\end{bmatrix} + i\sin(\delta)\begin{bmatrix}\cos(2\alpha_1)\\ \sin(2\alpha_1)\end{bmatrix}\right\} = \sqrt{\gamma}e^{i\Phi}\begin{bmatrix}\cos(\psi) & -\sin(\psi)\\ \sin(\psi) & \cos(\psi)\end{bmatrix}\begin{bmatrix}\cos(\chi)\\ i\sin(\chi)\end{bmatrix} \quad (36)$$

where $C_0$ is the numerically determined constant of proportionality in $Q = C_0/\delta_0^2$. Equating the real and imaginary parts of each polarization component gives:

$$\begin{aligned}
\sqrt{\gamma}\left[\cos(\Phi)\cos(\psi)\cos(\chi) + \sin(\Phi)\sin(\psi)\sin(\chi)\right] &= \sqrt{C_0}\delta_0 \cos(\delta)\cos(2\alpha_1)\\
\sqrt{\gamma}\left[\cos(\Phi)\sin(\psi)\cos(\chi) - \sin(\Phi)\cos(\psi)\sin(\chi)\right] &= \sqrt{C_0}\delta_0 \cos(\delta)\sin(2\alpha_1)\\
\sqrt{\gamma}\left[\sin(\Phi)\cos(\psi)\cos(\chi) - \cos(\Phi)\sin(\psi)\sin(\chi)\right] &= \sqrt{C_0}\delta_0 \sin(\delta)\cos(2\alpha_2)\\
\sqrt{\gamma}\left[\sin(\Phi)\sin(\psi)\cos(\chi) + \cos(\Phi)\cos(\psi)\sin(\chi)\right] &= \sqrt{C_0}\delta_0 \sin(\delta)\sin(2\alpha_2)
\end{aligned} \quad (37)$$

Solving for $(\delta_0, \delta, \alpha_1, \alpha_2)$ given a desired $(Q, \Phi, \psi, \chi)$ prescribes

$$\begin{aligned}
\delta_0 &= \sqrt{Q/C_0}\\
\delta &= \tan^{-1}\left\{\frac{\sqrt{1-\cos(2\Phi)\cos(2\chi)}}{\sqrt{1+\cos(2\Phi)\cos(2\chi)}}\right\}\\
\alpha_1 &= \frac{1}{2}\tan^{-1}\left\{\frac{\cos(\Phi)\sin(\psi)\cos(\chi) - \sin(\Phi)\cos(\psi)\sin(\chi)}{\cos(\Phi)\cos(\psi)\cos(\chi) + \sin(\Phi)\sin(\psi)\sin(\chi)}\right\}\\
\alpha_2 &= \frac{1}{2}\tan^{-1}\left\{\frac{\sin(\Phi)\sin(\psi)\cos(\chi) + \cos(\Phi)\cos(\psi)\sin(\chi)}{\sin(\Phi)\cos(\psi)\cos(\chi) - \cos(\Phi)\sin(\psi)\sin(\chi)}\right\}
\end{aligned} \quad (38)$$

Therefore, TCMT predicts a complete mapping of any desired Q-factor, polarization and phase to the four geometric parameters tuned by the metasurface. Note, in Eqn. (38) $\delta$ is bound in the range $[0, \pi/2]$, and when the arctangent function is implemented as atan2, $\alpha_1$ and $\alpha_2$ are bound in the range $[-\pi/2, \pi/2]$; however, due to the symmetry of an ellipse, there are other choices of bounds for these three angles that equivalently span the space. For instance, we may alternatively define $\delta$ across all four quadrants with appropriate rotations of either or both $\alpha_1, \alpha_2$ by 90°. E.g., $\delta = -3\pi/4, \alpha_1 = 0, \alpha_2 = 0$ gives an identical structure to $\delta = \pi/4, \alpha_1 = \pi/2, \alpha_2 = \pi/2$. Hence, for the LP case of the main text, we may keep $\alpha_1, \alpha_2$ constant and vary $\delta$ across $2\pi$ to implement the phase gradient, allowing the interpretation that $\Phi = \delta$ in that case.

To confirm this mapping, we perform a series of FDTD simulations. We note that the dependence of Q-factor on perturbation strength is well-established for q-BICs, and indeed is a defining feature; it is true by construction. We therefore limit this 4D parameter sweep to a 3D parameter sweep by keeping $\delta_0 = 130nm$ and varying $\delta, \alpha_1, \alpha_2$. This makes constructing such a library computationally tractable. We study $13 \times 19 \times 19 = 4693$ geometries with full-wave simulations with the values:

$$\begin{aligned} \delta &= linspace(0, 180°, 13) \\ \alpha_1 &= linspace(0, 90°, 19) \\ \alpha_2 &= linspace(0, 90°, 19) \end{aligned} \quad (39)$$

By symmetry, we need not simulate the full ranges, allowing us to exclude

$$\begin{aligned} \delta &= (180°, 360°] \\ \alpha_1 &= (90°, 180°] \\ \alpha_2 &= (90°, 180°] \end{aligned} \quad (40)$$

which may be recovered by symmetry. This enables a further speed up of computation.

Figure S2(a,b) show the top and side views of the geometries studied, where we have chosen the specific values $H = 500nm$, $D_0 = 190nm$, $a = 400nm$, and $\delta_0 = 130nm$, and study light incident from the substrate. For each of the 4693 geometries, we simulate the response due to incident x polarized light and y polarized light, making a total of 9386 simulations. This allows easy re-simulation of any incident polarization by appropriate complex weighting the x and y polarized data. To analyze the results, we (1) determine the resonant frequency in each case, (2) determine the maximally resonant polarization state by sweeping $\psi, \chi$, and then (3) record the reflection phase $\Phi$ at the resonant wavelength when that polarization state is incident. This methodology gives the full-wave results for $\Phi, \psi, \chi$ for each chosen $\delta, \alpha_1, \alpha_2$, which we may compare to the equivalent, but much more easily computed, predictions of TCMT.

To visualize the 3D library, Fig. S2(c,d) depict the retrieved values $\Phi, 2\psi, 2\chi$ by depicting 12 colormaps for each (each map corresponding to a fixed value of $\delta$) in which $\alpha_1$ varies along the x axis and $\alpha_2$ varies along the y axis [see the inset of Fig. S2(d)]. We see excellent agreement between the FDTD simulations and the predictions of TCMT, confirming complete control of the scattering of the q-BIC.

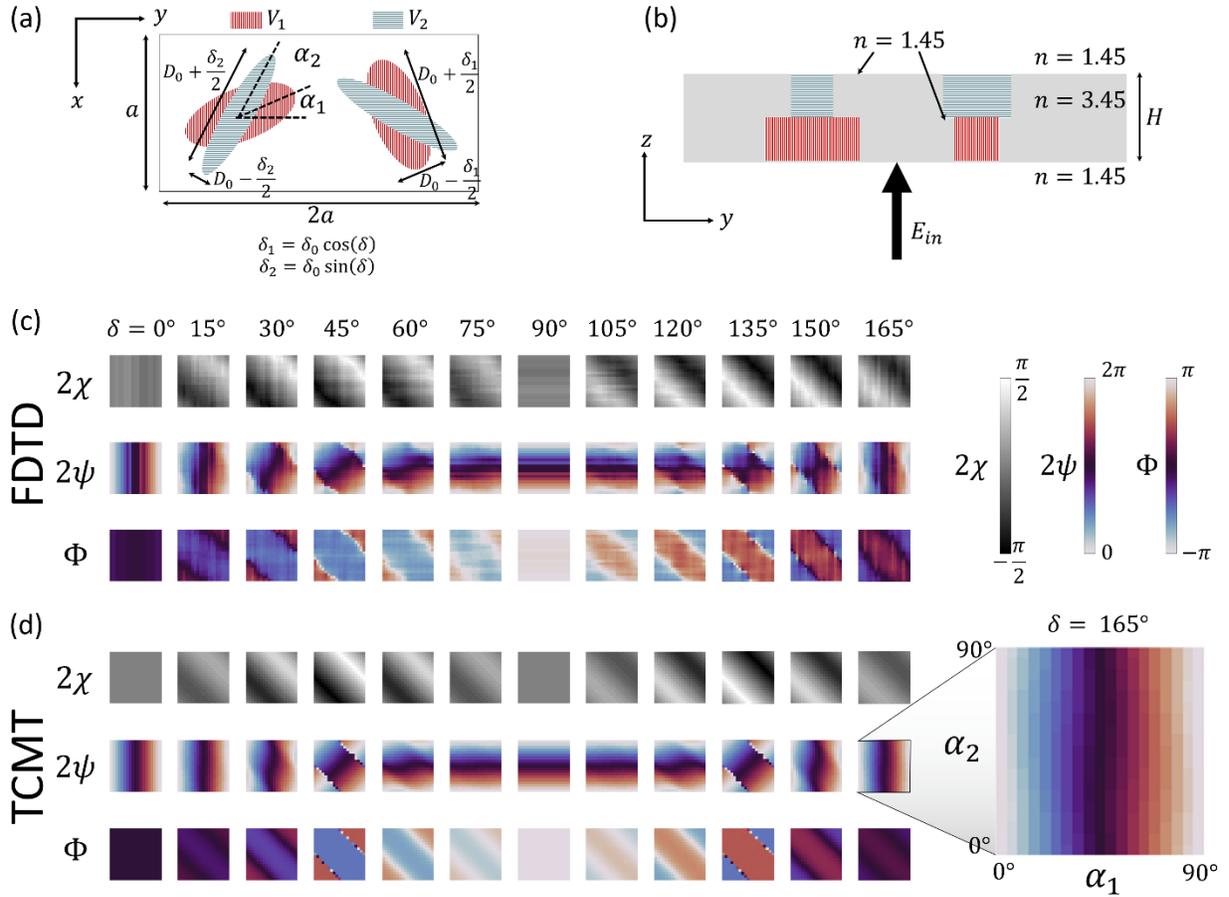

**Figure S2.** (a,b) Top and side view of the structures under study, with geometric parameters $\delta_0, \delta, \alpha_1, \alpha_2$ defined. Here we study devices with $H = 500nm$, $D_0 = 190nm$, $a = 400nm$, and $\delta_0 = 130nm$. (c) FDTD and (d) TCMT results for the Poincare latitude, $2\chi$, and Poincare longitude, $2\psi$, and reflection phase, $\Phi$ of the resonantly reflecting state for choice values of $\delta, \alpha_1, \alpha_2$. These results are for light incident from the substrate.

Finally, we may also invert the relations in Eqn. (38) to obtain $(Q, \Phi, \psi, \chi)$ given by a specific $(\delta_0, \delta, \alpha_1, \alpha_2)$. This, in turn, clarifies the mapping of the geometrical parameters $(\delta, \alpha_1, \alpha_2)$ to some set $(\alpha', \beta', \gamma')$ appropriate for a target EP state. We find:

$$\begin{aligned}
\Phi &= \alpha' = \operatorname{atan2}(Y_1, X_1) \\
\psi &= \beta' = \operatorname{atan2}(Y_2, X_2) \\
\chi &= \gamma' = \frac{\pi}{2} + \operatorname{atan2}(Y_3, X_3)
\end{aligned} \qquad (41)$$

where

$$\begin{aligned}
Y_1 &= B - 2\cos(2\delta) \\
X_1 &= 2\sin(2\delta)\cos(2\alpha_1 - 2\alpha_2) \\
Y_2 &= \cos(2\alpha_1)\left(B - 2\cos(\delta)^2\right) - 2\cos(4\alpha_2 - 2\alpha_1)\sin(\delta)^2 \\
X_2 &= \sin(2\alpha_1)\left(B - 2\cos(\delta)^2\right) - 2\sin(4\alpha_2 - 2\alpha_1)\sin(\delta)^2 \\
Y_3 &= -\sqrt{4 + 2B} \\
X_3 &= \frac{(B-2)\sqrt{2+B}}{\sqrt{2}\sin(2\delta)\sin(2\alpha_1 - 2\alpha_2)} \\
B &= \sqrt{3 + \cos(4\alpha_1 - 4\alpha_2) + 2\cos(4\delta)\sin(2\alpha_1 - 2\alpha_2)}
\end{aligned} \qquad (42)$$

We note that these relations are ill-conditioned near CP states, where $\Phi$ and $\Psi$ are indistinguishable. The CP case was treated in Ref. [37] of the main text.

## S4. Details of device geometries used in the main text

Here, we provide a table of the exact dimensions of the ellipses used for the full-wave simulations in the main text, and in the experiments. Figure S3 reports each nonlocal phase gradient, and the marked meta-units show the dimensions for the four meta-units highlighted in Figure 2(d). The elliptical features are made of a lossless dielectric with $n=1.45$, embedded in a thin film of lossless dielectric with $n=3.45$. The thickness of each ellipse layer is $250nm$ to make a total thickness of this thin film is $H=500nm$. The pitch of the array (i.e., center-to-center spacing of each feature) is $a=400nm$. The substrate and superstrate are both lossless dielectric with $n=1.45$.

We note that in these geometries we break from the assumptions in the previous section, in which $D_0, \delta_0$ were kept constant. In practice, both the resonant frequency and Q-factor drift slightly as a function of $\delta, \alpha_1, \alpha_2$. Leaving these drifts uncorrected results in inhomogeneous broadening for a device in which $\delta, \alpha_1, \alpha_2$ vary spatially, and so a library in which they are constant is desirable. Starting from the geometries in the previous section, we may adjust $D_0, \delta_0$ to re-obtain the desired resonant frequency and Q-factor. By increasing (decreasing) $D_0$ we may blueshift (redshift) the resonant frequency to the desired value (kept constant across the meta-units). And by increasing (decreasing) $\delta_0$ we may decrease (increase) the Q-factor.

Finally, Figure S4 shows the equivalent parameters for the experimental devices in Figure 5 of the main text. The other parameters not shown in Figure S4 are provided in the Method section in the main text. The photos of the fabricated devices are shown in Figure S6. Note that the field profile of the a-BIC used for the millimeter-wave devices has a single lobe in *z*-direction as in Figure S5 instead of double lobes in the all-dielectric devices of numerical demonstration (Fig. 2).

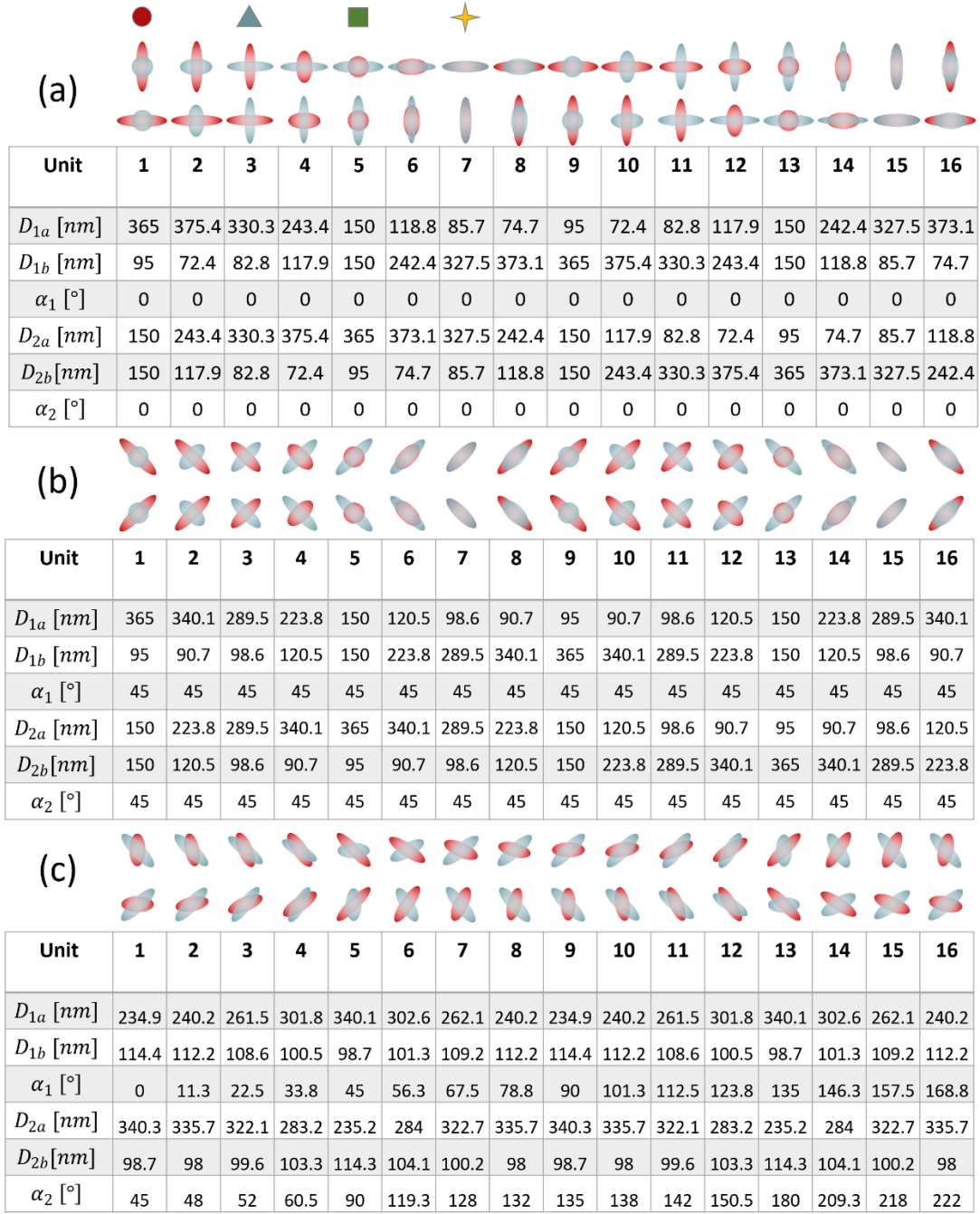

**Figure S3**. (a-c) Detailed geometric parameters of the nonlocal phase gradients in Figure 4(a-c) of the main text. The four marked meta-units in (a) are the four meta-units used for the phase profiles in Fig. 2(d) of the main text.

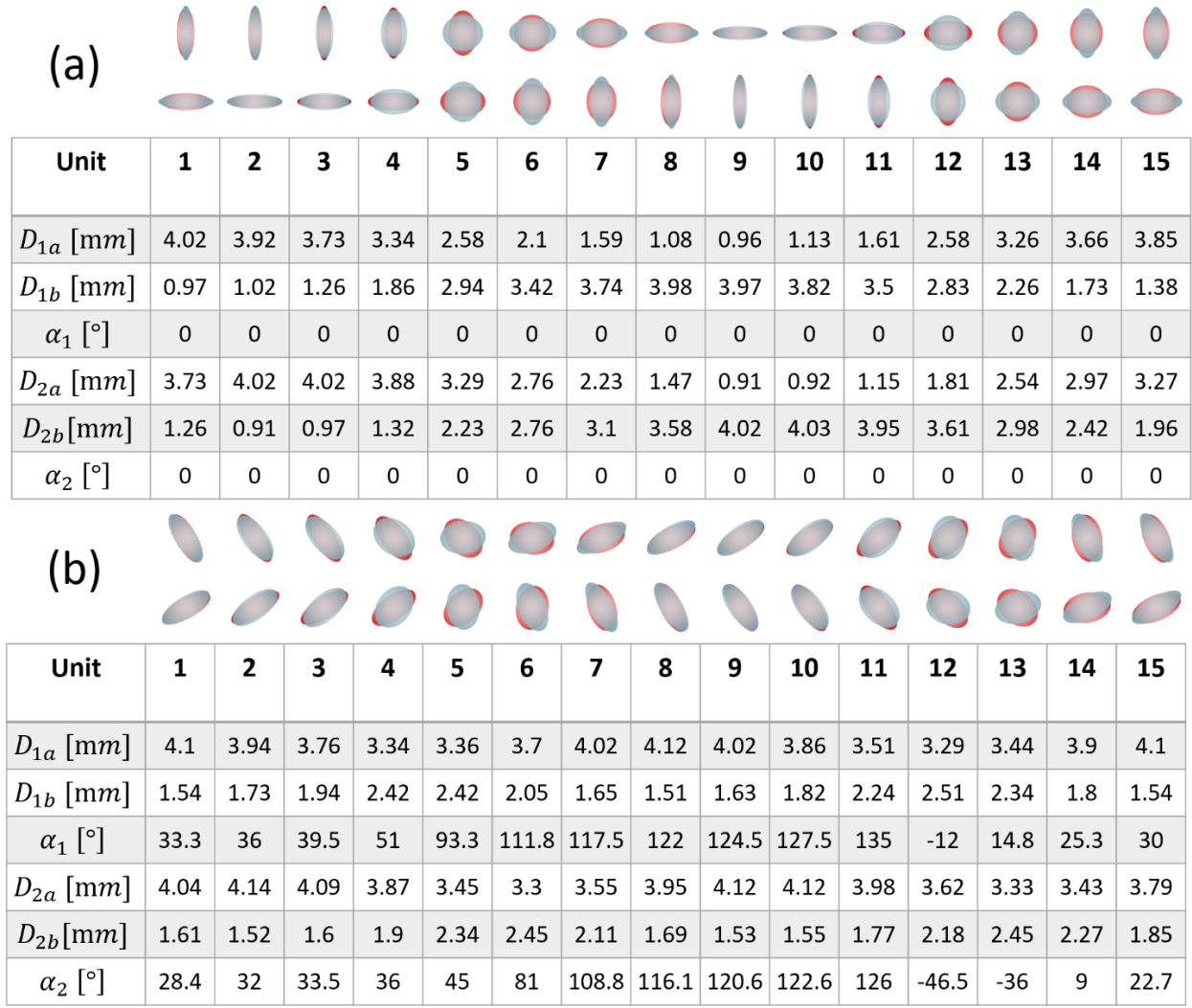

(a)

| Unit | 1 | 2 | 3 | 4 | 5 | 6 | 7 | 8 | 9 | 10 | 11 | 12 | 13 | 14 | 15 |
|---|---|---|---|---|---|---|---|---|---|---|---|---|---|---|---|
| $D_{1a}$ [mm] | 4.02 | 3.92 | 3.73 | 3.34 | 2.58 | 2.1 | 1.59 | 1.08 | 0.96 | 1.13 | 1.61 | 2.58 | 3.26 | 3.66 | 3.85 |
| $D_{1b}$ [mm] | 0.97 | 1.02 | 1.26 | 1.86 | 2.94 | 3.42 | 3.74 | 3.98 | 3.97 | 3.82 | 3.5 | 2.83 | 2.26 | 1.73 | 1.38 |
| $\alpha_1$ [°] | 0 | 0 | 0 | 0 | 0 | 0 | 0 | 0 | 0 | 0 | 0 | 0 | 0 | 0 | 0 |
| $D_{2a}$ [mm] | 3.73 | 4.02 | 4.02 | 3.88 | 3.29 | 2.76 | 2.23 | 1.47 | 0.91 | 0.92 | 1.15 | 1.81 | 2.54 | 2.97 | 3.27 |
| $D_{2b}$ [mm] | 1.26 | 0.91 | 0.97 | 1.32 | 2.23 | 2.76 | 3.1 | 3.58 | 4.02 | 4.03 | 3.95 | 3.61 | 2.98 | 2.42 | 1.96 |
| $\alpha_2$ [°] | 0 | 0 | 0 | 0 | 0 | 0 | 0 | 0 | 0 | 0 | 0 | 0 | 0 | 0 | 0 |

(b)

| Unit | 1 | 2 | 3 | 4 | 5 | 6 | 7 | 8 | 9 | 10 | 11 | 12 | 13 | 14 | 15 |
|---|---|---|---|---|---|---|---|---|---|---|---|---|---|---|---|
| $D_{1a}$ [mm] | 4.1 | 3.94 | 3.76 | 3.34 | 3.36 | 3.7 | 4.02 | 4.12 | 4.02 | 3.86 | 3.51 | 3.29 | 3.44 | 3.9 | 4.1 |
| $D_{1b}$ [mm] | 1.54 | 1.73 | 1.94 | 2.42 | 2.42 | 2.05 | 1.65 | 1.51 | 1.63 | 1.82 | 2.24 | 2.51 | 2.34 | 1.8 | 1.54 |
| $\alpha_1$ [°] | 33.3 | 36 | 39.5 | 51 | 93.3 | 111.8 | 117.5 | 122 | 124.5 | 127.5 | 135 | -12 | 14.8 | 25.3 | 30 |
| $D_{2a}$ [mm] | 4.04 | 4.14 | 4.09 | 3.87 | 3.45 | 3.3 | 3.55 | 3.95 | 4.12 | 4.12 | 3.98 | 3.62 | 3.33 | 3.43 | 3.79 |
| $D_{2b}$ [mm] | 1.61 | 1.52 | 1.6 | 1.9 | 2.34 | 2.45 | 2.11 | 1.69 | 1.53 | 1.55 | 1.77 | 2.18 | 2.45 | 2.27 | 1.85 |
| $\alpha_2$ [°] | 28.4 | 32 | 33.5 | 36 | 45 | 81 | 108.8 | 116.1 | 120.6 | 122.6 | 126 | -46.5 | -36 | 9 | 22.7 |

**Figure S4**. (a-b) Detailed geometric parameters of the nonlocal phase gradients in Figure 5(d,e) of the main text.

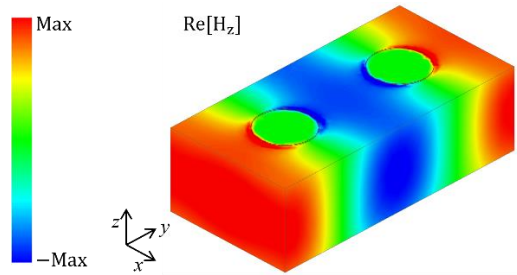

**Figure S5**. Field profile of the a-BIC for millimeter-wave devices.

## S5. Transmission measurement

The photo of the experiential setup for the transmission measurement introduced in the Method section in the main text is shown in Figure S7(a). We show supplemental experimental results about the transmission measurement in this section.

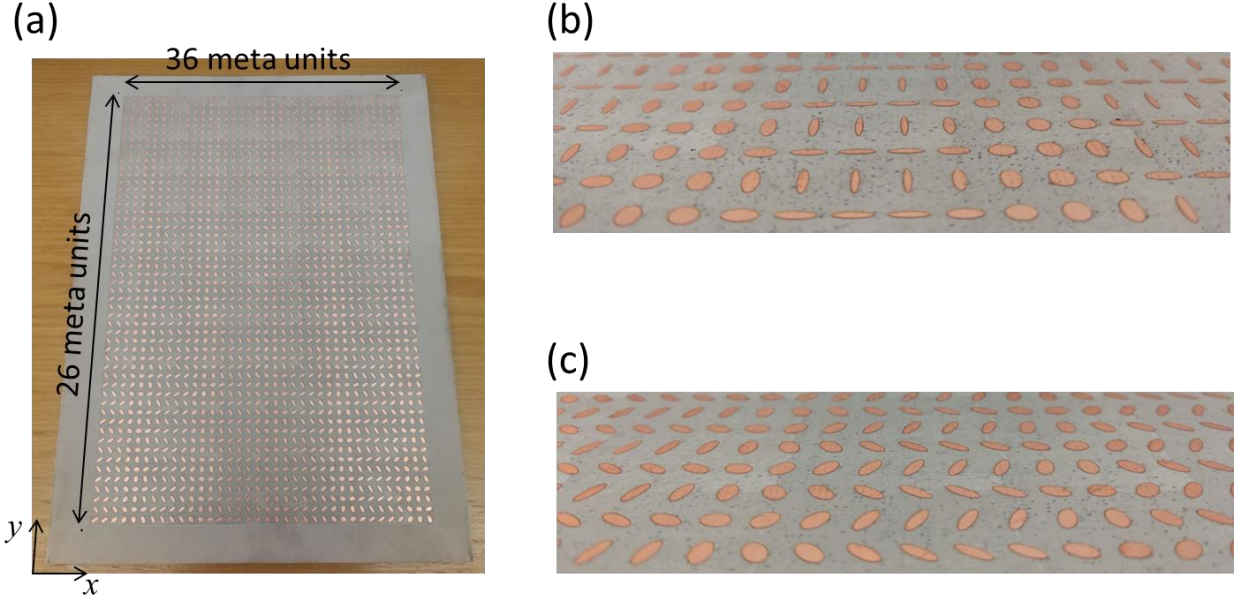

**Figure S6**. (a) Photo of the whole device of the metasurface with elliptic polarization state. (b) Nonlocal metasurface with linear polarization state and (c) the metasurface with elliptic polarization state.

The resonant polarization states were identified by measuring the *S*-parameters for normally incident excitation. We evaluated the transmitted power using $\langle J'_v | J'_v \rangle$, where $|J'_v\rangle = S^{(t)} |J_v\rangle$, for the input wave of the arbitrary polarization. The Jones vector $|J_v\rangle$ is calculated for the given elliptical parameter $(\psi, \chi)$ and $S^{(t)}$ is the *S*-parameters in the transmission measurement (superscript $(t)$), which are defined as $S^{(t)} = \begin{pmatrix} S^{(t)}_{HH} & S^{(t)}_{HV} \\ S^{(t)}_{VH} & S^{(t)}_{VV} \end{pmatrix}$. The subscripts, *H* and *V,* stand for horizontal and vertical polarizations, respectively (e.g., $S^{(t)}_{HV}$ is the *S*-parameter from the vertical polarization to the horizontal polarization). The resonant polarization state corresponds to the transmission minimum in Figure S7 (b-c), showing $\psi = 1°$ and $\chi = 1°$ for the nonlocal metasurface with the linear polarization and $\psi = 62°$ and $\chi = -24°$ for that with the elliptical polarization. These resonant polarizations agree well with designed polarizations: $\psi = 0°$ and $\chi = 0°$ for the linear polarization and $\psi = 60°$ and $\chi = -15°$ for the elliptical polarization.

In the main text, we show in Fig. 5(f-g) the transmission of resonant polarization, $|\langle e_1|S^{(t)}|e_1\rangle|^2$, and orthogonal polarization, $|\langle e_2|S^{(t)}|e_2\rangle|^2$, where $|e_1\rangle$ is the resonant polarization state of a nonlocal metasurface and $|e_2\rangle$ is the polarization orthogonal to $|e_1\rangle$. We show the remaining components in Figures S7(e-f) and S7(i-j). Figures S7(d,g,h,k) are already shown in Fig. 5 in the main text. We used $|e_1\rangle = \begin{pmatrix} 1 \\ 0 \end{pmatrix}$ and $|e_2\rangle = \begin{pmatrix} 0 \\ 1 \end{pmatrix}$ for the nonlocal metasurface with linear polarization, while $|e_1\rangle = \begin{pmatrix} 0.53 \\ 0.70 + 0.47i \end{pmatrix}$ and $|e_2\rangle = \begin{pmatrix} 0.85 \\ -0.44 - 0.30i \end{pmatrix}$ for that with elliptical polarization; $|e_1\rangle$ of each is calculated by the elliptical parameter $(\psi, \chi) = (0°, 0°)$ and $(\psi, \chi) = (60°, -15°)$, respectively.

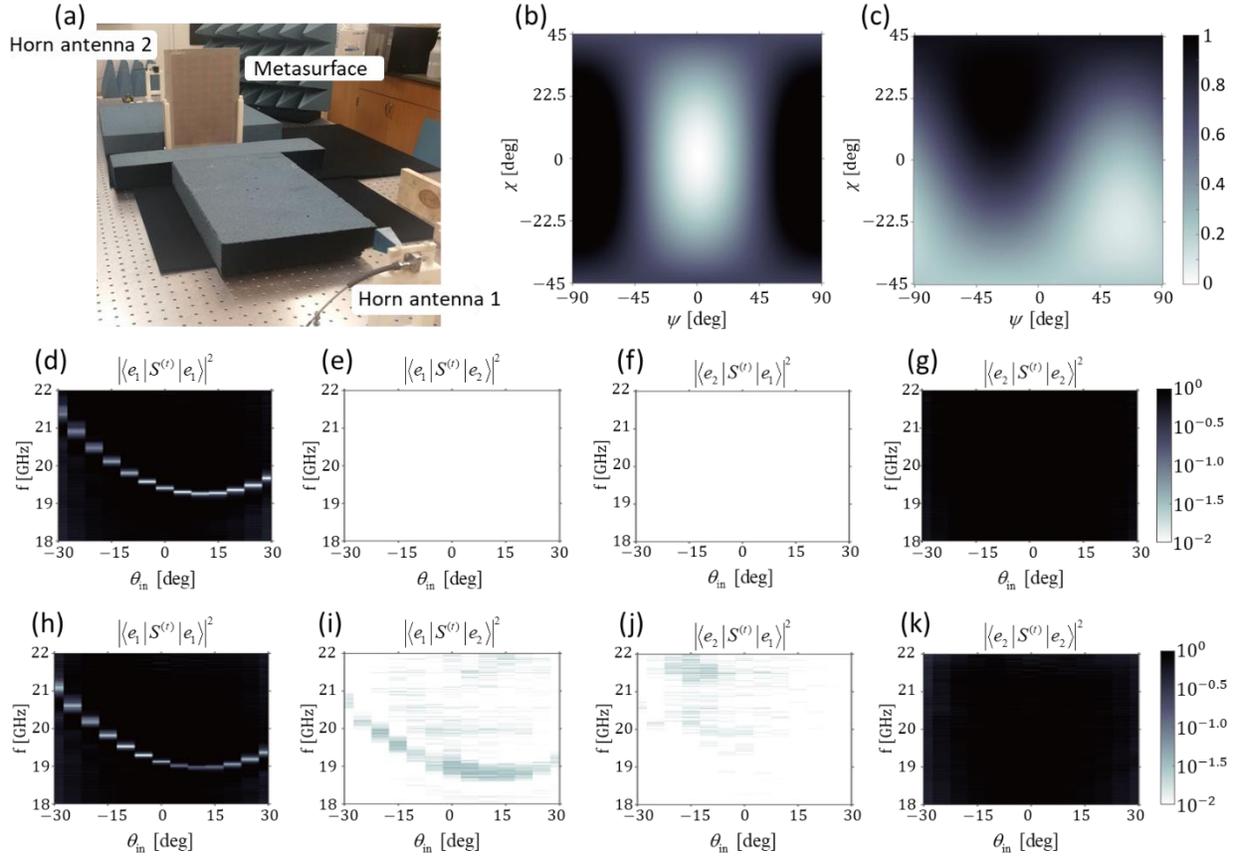

**Figure S7**. (a) Photo of the experimental setup for the transmission measurement. (b) Transmission as a function of input polarization for nonlocal metasurfaces with linear polarization ($\psi = 0°$ and $\chi = 0°$) and (b) elliptic polarization ($\psi = 60°$ and $\chi = -15$). (d-g) Transmission spectra as a function of incident angle, $\theta_{\text{in}}$, for nonlocal metasurfaces with linear polarization and (h-k) elliptic polarization.

## S6. Reflective measurement

The photo of the experiential setup of the reflection measurement introduced in the Method is shown in Figure S8(a). We show supplemental experimental results about the reflection measurement in this section.

Same as in the section S6, we complete the matrix element, not shown in Fig. 5 (h-i) of the main text. $\left|\langle e^*_1|\tilde{S}^{(r)}|e_1\rangle\right|^2$ and $\left|\langle e^*_2|\tilde{S}^{(r)}|e_2\rangle\right|^2$ are already shown in the main text. In addition

to them, all matrix components are shown in Figures S8(b-e) for the nonlocal metasurface with linear polarization and (f-i) for that with elliptical polarization state. $\tilde{S}^{(r)}$ are the S-parameters in the momentum space for reflected light (superscript $(r)$), defined as $\tilde{S}^{(r)} = \begin{pmatrix} \tilde{S}^{(r)}_{HH} & \tilde{S}^{(r)}_{HV} \\ \tilde{S}^{(r)}_{VH} & \tilde{S}^{(r)}_{VV} \end{pmatrix}$. The polarization basis $|e_1\rangle$ and $|e_2\rangle$ are the same as those used in the transmission results.

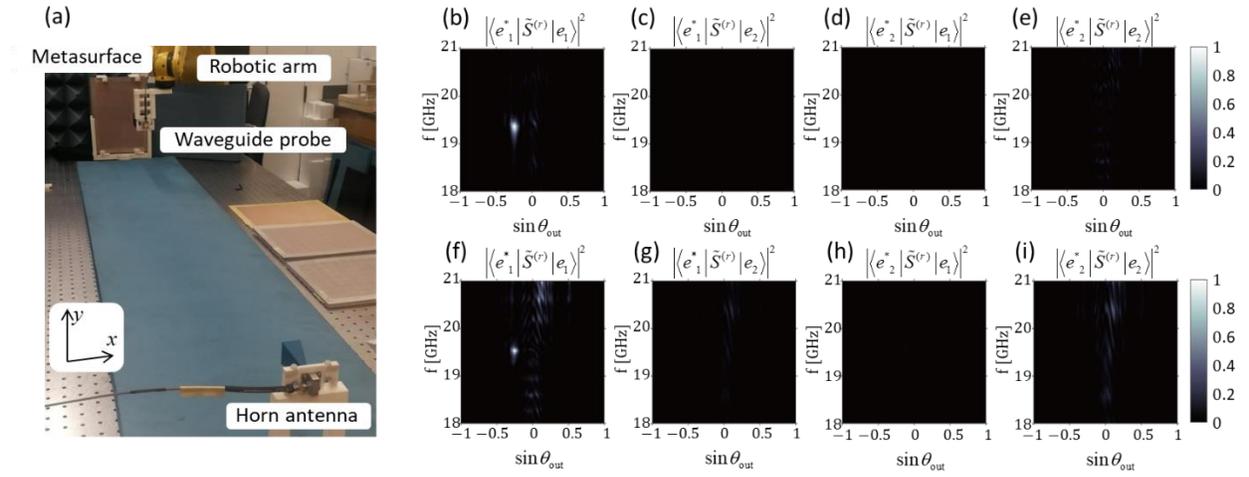

**Figure S8**. (a) Photo of the experimental setup for the reflection measurement. (b-e) Reflection spectra as a function of reflective angle, $\theta_{out}$, for nonlocal metasurfaces with linear polarization and (f-i) elliptical polarization.